\documentclass{SciPost}

\hypersetup{
    colorlinks,
    linkcolor={red!50!black},
    citecolor={blue!50!black},
    urlcolor={blue!80!black}
}

\usepackage[bitstream-charter]{mathdesign}
\DeclareSymbolFont{usualmathcal}{OMS}{cmsy}{m}{n}
\DeclareSymbolFontAlphabet{\mathcal}{usualmathcal}

\fancypagestyle{SPstyle}{
\fancyhf{}
\lhead{\colorbox{scipostblue}{\bf \color{white} ~SciPost Physics }}
\rhead{{\bf \color{scipostdeepblue} ~Submission }}

\fancyfoot[C]{\textbf{\thepage}}
}
\usepackage{booktabs}

\def\CF{\mathrm{C_F}}
\def\CFsq{\mathrm{C_F^2}}
\def\CFcub{\mathrm{C_F^3}}
\def\CFfour{\mathrm{C_F^4}}

\def\CA{\mathrm{C_A}}
\def\CAsq{\mathrm{C_A^2}}
\def\CAcub{\mathrm{C_A^3}}

\def\W{\mathcal{W}}

\def\A{\mathcal{A}} 
\def\B{\mathcal{B}} 
\def\C{\mathcal{C}}
\def\F{\mathcal{F}}

\def\cS{\mathcal{S}}
\def\cR{\mathcal{R}}
\def\G{\mathcal{G}}

\def\s{\sigma}
\def\S{\Sigma}
\renewcommand{\O}{\Omega}
\newcommand{\Ob}{\bar{\Omega}}
\def\inn{\mathrm{in}}
\def\out{\mathrm{out}}
\def\X{{\scriptscriptstyle X}}
\def\R{{\scriptscriptstyle\mathrm{R}}}
\def\V{{\scriptscriptstyle\mathrm{V}}}
\def\L{{\scriptscriptstyle\mathrm{L}}}
\def\P{{\scriptscriptstyle\mathrm{P}}}

\renewcommand{\d}{\mathrm{d}}
\newcommand{\as}{\alpha_s}
\newcommand{\asb}{\bar{\alpha}_s}

\def\ktt{\scriptscriptstyle\mathrm{k_t}}
\def\ca{\scriptscriptstyle\mathrm{C/A}}

\def\ng{\text{ng}}
\def\cl{\text{cl}}
\def\MC{\scriptscriptstyle\mathrm{MC}}

\def\Uh{\hat{\mathcal{U}}}
\def\uh{\hat{u}}
\begin{document}

\pagestyle{SPstyle}

\begin{center}{\Large \textbf{\color{scipostdeepblue}{
Structure of non-global logarithms with Cambridge/Aachen clustering\\
}}}\end{center}

\begin{center}\textbf{
K. Khelifa-Kerfa\textsuperscript{1,2,3,4$\star$}
}\end{center}

\begin{center}
{\bf 1} Department of Physics, Faculty of Science and Technology, Relizane University, Relizane 48000, Algeria
\\
{\bf 2} Laboratory of Physics of Experimental Techniques and Applications, University of Médéa, Médéa 26000, Algeria
\\
{\bf 3} Laboratory of Thin Layers and Advanced Technologies, Relizane University, Relizane 48000, Algeria
\\
{\bf 4} Laboratory of Mathematics and Applications, University of Chlef, Chlef 02000, Algeria
\\[\baselineskip]
$\star$ \href{mailto:kamel.khelifakerfa@univ-relizane.dz}{\small kamel.khelifakerfa@univ-relizane.dz}\,
\end{center}

\section*{\color{scipostdeepblue}{Abstract}}
\textbf{\boldmath{%
%
We determine the structure of both Abelian and non-Abelian non-global logarithms up to four loops for $e^+e^-$ processes in perturbative QCD, where final-state jets are defined using the Cambridge--Aachen (C/A) clustering algorithm. The calculations are performed within the soft (eikonal) approximation using strong-energy ordering of the final-state partons for the case of the dijet invariant mass. The resulting expressions include full colour and complete jet-radius dependence. Compared to the anti-$k_t$ and $k_t$ clustering algorithms, the C/A distribution minimises the impact of these non-global logarithms, making it the preferred choice among the three algorithms.
}}

\vspace{\baselineskip}

\noindent\textcolor{white!90!black}{%
\fbox{\parbox{0.975\linewidth}{%
\textcolor{white!40!black}{\begin{tabular}{lr}%
  \begin{minipage}{0.6\textwidth}%
    {\small Copyright attribution to authors. \newline
    This work is a submission to SciPost Physics. \newline
    License information to appear upon publication. \newline
    Publication information to appear upon publication.}
  \end{minipage} & \begin{minipage}{0.4\textwidth}
    {\small Received Date \newline Accepted Date \newline Published Date}%
  \end{minipage}
\end{tabular}}
}}
}


\vspace{10pt}
\noindent\rule{\textwidth}{1pt}
\tableofcontents
\noindent\rule{\textwidth}{1pt}
\vspace{10pt}

\section{Introduction}
\label{sec:intro}

Precision studies of QCD radiation at colliders increasingly rely on observables that are sensitive to restricted regions of phase space. Among these, jet substructure observables such as the jet mass constitute paradigmatic examples of so-called non-global observables. Their perturbative distributions receive leading enhancements not only from soft and collinear radiation directly associated with the hard initiating parton, but also from correlated emissions that populate disparate angular regions. The latter give rise to non-global logarithms (NGLs), which are intrinsically entangled with the pattern of colour correlations in the event and, importantly, are highly sensitive to the detailed definition of the jet employed in the analysis \cite{Dasgupta:2001sh, Banfi:2002hw, Appleby:2002ke, Banfi:2010pa}.

The dependence of non-global observables on the jet algorithm manifests itself through two distinct mechanisms. First, the clustering sequence of the algorithm modifies the phase space available to primary (or ``global'') soft–collinear radiation and generates a tower of clustering logarithms (CLs) \cite{Banfi:2005gj, Banfi:2010pa, Kerfa:2012yae}. Second, the algorithm alters the survival probability of correlated secondary emissions that are the source of NGLs. These two effects — the modification of primary emission contributions (CLs) and the reshaping of secondary emission patterns (NGLs) — need therefore be treated within a unified framework to achieve a controlled resummation, at least at single-logarithmic accuracy.

Over the past two decades, substantial progress has been made for specific jet definitions. For the anti-$k_t$ algorithm \cite{Cacciari:2008gp}, whose rigid cone-like clustering simplifies the geometry of final-state jets, the resummation of NGLs has been achieved through various approaches. These include, in the large-$N_c$ limit, the Monte Carlo (MC) code of Dasgupta and Salam \cite{Dasgupta:2001sh} and the Banfi--Marchesini--Smye (BMS) integro-differential equation \cite{Banfi:2002hw}, along with finite-$N_c$ generalisations, both at the cross-section \cite{Weigert:2003mm, Hatta:2013iba, Hagiwara:2015bia, Hatta:2020wre} and amplitude levels \cite{Nagy:2007ty, Nagy:2017ggp, Forshaw:2025fif, Forshaw:2025bmo}, although some of the latter have been implemented only for a limited set of observables.
The current state-of-the-art for these resummations reaches next-to-leading NGLs accuracy, in the large-$N_c$ limit,  \cite{Banfi:2021owj, Banfi:2021xzn, Becher:2023vrh, FerrarioRavasio:2023kyg}.

For the $k_t$ algorithm \cite{Catani:1993hr, Ellis:1993tq}, only MC codes in the large-$N_c$ limit are currently available: the MC of \cite{Dasgupta:2001sh}, which handles both double-logarithmic (DL) and single-logarithmic (SL) observables, and the recently developed MC of \cite{Becher:2023znt} within the Soft-Collinear Effective Theory (SCET) framework, which currently treats only SL observables. The latter code also applies to the Cambridge-Aachen algorithm \cite{Dokshitzer:1997in, Wobisch:1998wt}. To date, the state-of-the-art resummation for these algorithms remains at leading NGLs (and CLs) in the large-$N_c$ limit. No next-to-leading NGLs (or CLs) or finite-$N_c$ resummations exist in the literature for $k_t$ and/or C/A large logarithms. We intend to address these challenges in future work.

Despite these advances, clustering algorithms differ qualitatively in the way they order emissions and recombine particles and/or pseudo-jets. The Cambridge/Aachen algorithm, which clusters purely according to angular separation, presents particular challenges: its clustering does not respect a simple scale ordering and, hence, resummation frameworks based on an evolution in emission hardness are not readily available. Consequently, for DL-sensitive observables there is, to date, no compact evolution equation or general-purpose MC algorithm that achieves the same level of automation as for anti-$k_t$ or $k_t$; progress for C/A has therefore relied on ``brute-force'' fixed-order calculations and tailored numerical treatments for specific observables such as the jet mass \cite{Delenda:2012mm, Khelifa-Kerfa:2025cdn}. This absence of an ordering variable renders C/A an interesting, yet technically demanding, setup for the study of both CLs and NGLs.

To make analytical progress in this difficult setting it is customary to introduce a few approximations. Two particularly useful simplifications are the soft (eikonal) approximation, which retains only the leading soft singularities of the emission amplitudes, and the strong-energy (or strong-$k_t$) ordering of emissions, which reduces the combinatorial complexity of both the clustering sequence and the phase space. Within this simplified framework, one can derive semi-analytic expressions for fixed-order contributions to CLs and NGLs and, crucially, identify patterns that survive beyond these approximations. Fixed-order computations up to two loops have already exposed several robust features---the so-called boundary or ``edge'' effect whereby CLs and NGLs remain non-vanishing as the jet radius $R\to0$, the dominant role of gluon-initiated channels through larger colour factors, the similarity of the gross features across different QCD environments (lepton, lepton--hadron and hadron--hadron), and the general tendency of some clustering algorithms to reduce the size of NGLs relative to anti-$k_t$ (see, for instance, \cite{Appleby:2002ke, Banfi:2005gj, Delenda:2006nf, Banfi:2010pa, Khelifa-Kerfa:2011quw, Dasgupta:2012hg, Kerfa:2012yae, Ziani:2021dxr, Benslama:2023gys, Becher:2023znt, Khelifa-Kerfa:2024udm, Bouaziz:2022tik, Khelifa-Kerfa:2025jev}).

In this work we extend the analyses of \cite{Khelifa-Kerfa:2025cdn} and \cite{Khelifa-Kerfa:2024hwx} by computing CLs and NGLs for the C/A clustering algorithm to three and four loops for the jet mass in $e^+e^-$ annihilation into dijet processes. Working within the approximations described above, we perform the angular integrations while retaining the full colour and jet-radius ($R$) dependence, using high-precision numerical methods because closed-form expressions are not available. The higher-order results allow us to test exponentiation patterns and to motivate a next-to-leading-logarithmic (NLL) resummed form factor in which the global Sudakov factor multiplies exponentials that encode our fixed-order results for CLs and NGLs.

The way the C/A algorithm is structured contains, within it, the full $k_t$ algorithm, which in turn contains the anti-$k_t$ algorithm. This implies that the C/A algorithm explores the full set of clustering orderings: it covers the range of possible orderings and clusterings of the final-state particles, whereas $k_t$ and anti-$k_t$ realise only subsets, with anti-$k_t$ covering the smallest set. Hence, just as the $k_t$ calculations may be regarded as corrections to the anti-$k_t$ results \cite{Dasgupta:2001sh, Banfi:2010pa, Khelifa-Kerfa:2011quw, Kerfa:2012yae, Khelifa-Kerfa:2024dut, Khelifa-Kerfa:2024hwx}, the C/A calculations represent corrections to the $k_t$ ones \cite{Khelifa-Kerfa:2025cdn}. We find, through three and four loops, that these corrections are generally of opposite sign to the corresponding $k_t$ corrections. This leads to a reduction in the magnitude of both NGLs and CLs coefficients at these loop orders, and we anticipate that the trend may persist at higher orders.
The jet-mass distribution appears to exhibit a pattern of exponentiation, as observed for the $k_t$ clustering \cite{Khelifa-Kerfa:2024dut, Khelifa-Kerfa:2024hwx}. Comparisons between the all-orders NLL resummations for anti-$k_t$ and $k_t$ — for which we use the output of the MC of \cite{Dasgupta:2001sh} for NGLs and/or CLs — and our fixed-order-based exponential for C/A confirm that C/A has the strongest impact in minimising the importance of non-global logarithms. Furthermore, up to four loops the finite-$N_c$ corrections do not appear to be significant. The dependence on the jet radius is moderate in the range $[0,1]$.

The paper is organised as follows. In Section~\ref{sec:definitions} we introduce our notation, kinematics, and the definition of the jet algorithm under study. Section~\ref{sec:fixed-order} presents the fixed-order framework and the explicit calculations of CLs and NGLs at three, and four loops. We note that the two-loop results have been previously computed in the literature (see, e.g., \cite{Appleby:2003sj, Delenda:2006nf, Banfi:2010pa, Delenda:2012mm, Khelifa-Kerfa:2011quw, Kerfa:2012yae, Benslama:2023gys}). In Section~\ref{sec:all-order} we compare our fixed-order-based exponentiation with all-orders Monte Carlo resummations for the anti-$k_t$ and $k_t$ algorithms, and discuss the impact of NGLs and CLs arising from the C/A algorithm. Section~\ref{sec:conclusion} summarises our conclusions and outlines potential directions for future work.

\section{Observable and jet algorithm definitions}
\label{sec:definitions}

The majority of this section follows the approach detailed in Ref.~\cite{Khelifa-Kerfa:2024hwx}. We shall therefore only report the essential ideas and notation herein to ensure the paper's self-consistency.
As stated in the introduction, we consider the invariant squared mass of the two leading jets in $e^+ e^-$ annihilation processes in the threshold limit, which serves as an example to illustrate our main points. In particular, we consider the partonic process:
\begin{align}
 e^+ + e^- \to q(p_a) + \bar{q}(p_b) + g_1(k_1) + \cdots + g_n(k_n),
\end{align}
where the four-momenta of the final-state partons are defined by
\begin{align}\label{eq:Def:Momenta}
 p_a = \tfrac{Q}{2}(1,0,0,1), \qquad
 p_b = \tfrac{Q}{2}(1,0,0,-1), \qquad
 k_i = \omega_i(1,s_i \cos\phi_i, s_i \sin\phi_i, c_i),
\end{align}
with $c_i \equiv \cos \theta_i$ and $s_i \equiv \sin\theta_i$. Here, $Q$ is the hard scale of the process, $\omega_i$ is the energy of the $i$th soft gluon, and $\theta_i$ and $\phi_i$ are its polar and azimuthal angles, respectively. Our calculations are performed within the eikonal (soft) approximation, assuming massless partons, neglecting recoil effects, and imposing a strong ordering of the gluon energies: $Q \gg \omega_1 \gg \omega_2 \gg \cdots \gg \omega_n$. Whilst these approximations significantly simplify the calculations, they only guarantee single-logarithmic accuracy. As this paper presents a pioneering study of the C/A jet algorithm beyond two loops, it is sufficient to work at this level of accuracy, particularly given the substantial complexity involved. Future work will extend this to higher-order logarithmic contributions.


The observable we consider in this paper is the normalised invariant mass squared of the final-state dijets:
\begin{align}\label{eq:Def:Rho}
 \varrho = \varrho_\R + \varrho_\L,
\end{align}
where $\varrho_{\R(\L)}$ denotes the normalised invariant mass squared of the right (left) jet, initiated by the final-state quark (antiquark). Given the analogous expressions for both jets, we present only the explicit formula for $\varrho_\R$. To this end, we have \cite{Khelifa-Kerfa:2024hwx}
\begin{align}\label{eq:Def:Rho-R}
 \varrho_\R &= \frac{4\,m_\R^2}{Q^2} = \frac{4}{Q^2} \left(p_a + \sum_{i \in j_\R} k_i\right)^2 = \sum_{i\in j_\R} \varrho_{\R,i} + \mathcal{O}\left(\frac{\omega^2}{Q^2}\right),
\notag\\
\varrho_{\R,i} &= \frac{8 \left(p_a \cdot k_i\right)}{Q^2} = 2 x_i \left(1 - c_i\right),
\end{align}
where we have defined the energy fraction of the $i$th gluon as $x_i \equiv 2 \omega_i/Q$. The sum in the first line runs over all gluon emissions that are clustered, by the C/A algorithm, into the right jet $j_\R$. For the left jet, $j_\R$ is replaced by $j_\L$. Terms of order $(\omega/Q)^2$ and higher are neglected, consistent with the soft approximation.

\subsection{C/A jet algorithm}
\label{sec:jet-algo}

The C/A jet algorithm was first introduced in Refs.~\cite{Dokshitzer:1997in, Wobisch:1998wt}. It can be considered a member of the generalised $k_t$ family of algorithms defined in Ref.~\cite{Cacciari:2011ma}, which comprises a set of sequential recombination jet algorithms parametrised by a continuous variable $p$. For $e^+ e^-$ processes, these algorithms proceed as follows:
\begin{enumerate}
\item For an initial list of final-state particles, define the following two distance measures for each pair $(i,j)$:
\begin{align}\label{eq:Def:DistanceMeasure}
 d_{ij} = \min\left(E_i^{2p}, E_j^{2p} \right) \frac{1 - \cos\theta_{ij}}{1 - c_R}, \qquad
 d_{iB} = E_i^{2p},
\end{align}
where $c_R = \cos R$ with $R$ the jet-radius parameter, $E_i$ is the energy of the $i$th particle, and $\cos\theta_{ij} = c_i c_j + s_i s_j \cos\phi_{ij}$ with $\phi_{ij} \equiv \phi_i - \phi_j$. The quantity $d_{iB}$ represents the distance of particle $i$ to the beam direction.

\item Find the smallest distance amongst all $d_{ij}$ and $d_{iB}$. If the minimum is a $d_{ij}$, then particles $i$ and $j$ are merged into a single pseudo-jet, with a momentum given by the sum of their constituent momenta (the E-scheme). If the minimum is a $d_{iB}$, then particle $i$ is identified as a final jet and removed from the list of particles.

\item Repeat steps 1 and 2 until no particles remain.
\end{enumerate}
For the C/A algorithm, the parameter $p = 0$. Consequently, the distance $d_{ij}$ is purely geometrical and $d_{iB} = 1$. A sufficient condition for $d_{ij}$ to be the smallest distance---and hence for the particle pair $(i, j)$ to be clustered---is simply
\begin{align}\label{eq:Def:ClusCond}
 1 -\cos\theta_{ij} < 1 - c_R.
\end{align}
In the small-angle limit, this condition reduces to $\theta_{ij} < R$. That is, two particles are merged if they lie within a circle of radius $R$ in the $(\theta,\phi)$ plane.

A pivotal aspect where the C/A algorithm differs from the anti-$k_t$ and $k_t$ algorithms in our analytical calculations is that we must consider all possible orderings of the distances $d_{ij}$ at a given loop order. For the anti-$k_t$ and $k_t$ algorithms, the strong energy ordering of the soft gluons automatically induces a unique hierarchy of the distances $d_{ij}$. The unique clustering sequence for the $k_t$ algorithm is, in fact, just one of the many possible sequences for the C/A algorithm. Therefore, the C/A distribution inherently contains the $k_t$ result. Furthermore, since the $k_t$ distribution itself contains the anti-$k_t$ result \cite{Banfi:2010pa, Khelifa-Kerfa:2011quw, Delenda:2012mm, Khelifa-Kerfa:2024hwx}, the C/A distribution encompasses both the anti-$k_t$ and $k_t$ results, constituting a much more general distribution.

For each possible ordering of the C/A distances $d_{ij}$, we apply the remaining steps of the algorithm to every gluon configuration at a given loop order (as will be detailed later). This procedure leads to a reshuffling of soft gluons in and out of the measured jets. We retain only those configurations where a mis-cancellation occurs between real emissions and virtual corrections. These are the configurations that give rise to the large logarithmic terms. Given the two types of gluon emissions---primary (Abelian) and secondary correlated (non-Abelian)---two distinct types of large logarithms arise, as noted in the introduction: (Abelian) clustering logarithms (CLs) and (non-Abelian) non-global logarithms (NGLs).

\subsection{Observable distribution}

Following our previous work \cite{Khelifa-Kerfa:2024hwx}, we consider the integrated cross-section for the normalised dijet mass squared, $\S(\rho)$, defined by:
\begin{align}\label{eq:Def:Xsec}
 \S(\rho) = \int \frac{1}{\s_0} \frac{\d\s}{\d\varrho} \Theta\left[\rho - \varrho\left(k_1, k_2, \dots , k_n\right)\right] \Xi\left(k_1, k_2, \dots,k_n\right) \d\varrho,
\end{align}
where $\rho$ is a veto on the (normalised) dijet mass squared $\varrho(k_1,k_2,\dots, k_n)$ defined in eq. \eqref{eq:Def:Rho}, and $\s_0$ is the corresponding Born cross-section. The function $\Xi(k_1,k_2, \dots,k_n)$ represents the C/A clustering function, which constrains the integration to the regions of phase space corresponding to mis-cancellations between real emissions and virtual corrections that contribute to the dijet mass.

The perturbative expansion of the jet-mass cross-section may be written as:
\begin{align}\label{eq:Def:Xsec-PTExpnasion}
 \S(\rho) = \S_1(\rho) + \S_2(\rho) + \cdots,
\end{align}
where the $m$th-order contribution is given by:
\begin{align}\label{eq:Def:Xsec-mOrder}
 \S_m(\rho) = \sum_\X \int_{x_1>x_2>\dots>x_m} \prod_{i=1}^m \d\Phi_i\, \Uh_m \W_{12 \dots m}^{\X}\, \Xi_m\left(k_1,k_2,\dots,k_m\right).
\end{align}
The various terms are defined as follows:
\begin{itemize}
\item The sum is over all possible gluonic configurations $\X$. Each gluon can be either real (R) or virtual (V), and $\X$ represents the set of all possible real/virtual assignments for the $m$ gluons at a given order. For instance, at one loop, $\X$ may be R or V, and at two loops, it may be RR, RV, VR, or VV. The number of distinct configurations at $m$th order in perturbation theory is $2^m$.

\item The integration is performed under the constraint of strong-energy ordering of the soft gluons, represented by $x_1 > x_2 > \dots > x_m$. This constraint may sometimes be relaxed, in which case a factor of $m!$ is included to avoid double-counting identical phase-space regions.

\item The phase-space element for the emission of the $i$th gluon is given by:
\begin{align}\label{eq:Def:PhaseSpceFactor}
 \d\Phi_i = \asb \, \frac{\d x_i}{x_i} \d c_i \, \frac{\d \phi_i}{2\pi},
\end{align}
with $\asb = \as/\pi$.

\item The measurement operator at order $m$, $\Uh_m$, first introduced in Ref.~\cite{Schwartz:2014wha}, selects gluonic configurations that contribute to the dijet mass observable with a value less than $\rho$. Configurations that do not contribute to the dijet mass, or that yield a value larger than $\rho$, are excluded from the integrated cross-section. It can be shown that this operator factorises into a product of single-gluon measurement operators \cite{Schwartz:2014wha, Khelifa-Kerfa:2015mma, Khelifa-Kerfa:2024roc, Khelifa-Kerfa:2024hwx}:
\begin{align}\label{eq:Def:MeasOperator}
 \Uh_m = \prod_i \uh_i, \qquad
\uh_i = \Theta_i^\V + \Theta_i^\R \left[ \Theta_i^\out + \Theta_i^\inn \Theta(\rho - \varrho_i)  \right] = 1- \Theta_i^\R \Theta_i^\inn \Theta^\rho_i.
\end{align}
Here, the step functions $\Theta_i^{\R(\V)}$ equal 1 if gluon $i$ is real (virtual) and 0 otherwise; $\Theta_i^{\inn(\out)}$ equals 1 if gluon $i$ is inside (outside) either jet and 0 otherwise; and $\Theta^\rho_i$ equals 1 if $\varrho_i > \rho$ and 0 otherwise. The complementarity relations $\Theta_i^\R + \Theta_i^\V = 1$ and $\Theta_i^\inn + \Theta_i^\out = 1$ hold. The final equality in \eqref{eq:Def:MeasOperator} follows from these relations.

\item The eikonal amplitude squared for the emission of $m$ soft, energy-ordered gluons in configuration $\X$, normalised to the Born cross-section, is denoted $\W_{12 \dots m}^\X$. Detailed expressions for $e^+ e^-$ processes are provided in Ref.~\cite{Delenda:2015tbo, Benslama:2020wib} (and for hadronic processes with three hard legs in Ref.~\cite{Khelifa-Kerfa:2020nlc}), with explicit results given up to four loops.

\item The C/A clustering function at $m$th order, $\Xi_m(k_1,k_2, \dots, k_m)$, discussed in the previous section, encapsulates the clustering constraints imposed by the C/A algorithm on a specific configuration $\X$ for it to contribute to the dijet mass.
\end{itemize}
In the next section, we present detailed calculations of the third- and fourth-order terms in the perturbative expansion \eqref{eq:Def:Xsec-PTExpnasion}. First, however, we recapitulate the one- and two-loop results, which are identical to the $k_t$ algorithm case presented in Ref.~\cite{Khelifa-Kerfa:2024hwx}.

\section{Fixed-order calculations}
\label{sec:fixed-order}

From the definition of the generalised $k_t$ algorithm presented in the previous section, it is straightforward to see that all members of this family are identical at one-loop order. At two-loop order, the $k_t$ ($p=1$) and C/A ($p=0$) algorithms begin to differ from the anti-$k_t$ ($p=-1$), while remaining identical to each other. This occurs because, at two loops, we have the emission of two soft gluons $k_1$ and $k_2$, and the possible distance measures are simply $d_{1B}$, $d_{2B}$, and $d_{12}$. Gluonic configurations that lead to the appearance of CLs and NGLs typically correspond to the clustering of the two soft gluons first. This requires that $d_{12}$ be the smallest distance \cite{Banfi:2010pa, Khelifa-Kerfa:2011quw, Delenda:2012mm, Khelifa-Kerfa:2024hwx}. In analytical calculations, each gluon is typically assigned an initial position, either inside or outside the measured jets. The clustering of the two gluons means that the harder gluon $k_1$ pulls the softer gluon $k_2$ into its region, resulting in both gluons being either inside or outside the jets, depending on the initial location of $k_1$. For both the $k_t$ and C/A algorithms, this is the only possible ordering of the distance measures that gives rise to large logarithms not present in the anti-$k_t$ case. The difference between the two algorithms emerges at three loops, as will be discussed later.

\begin{figure}
\centering
\includegraphics[scale=0.53]{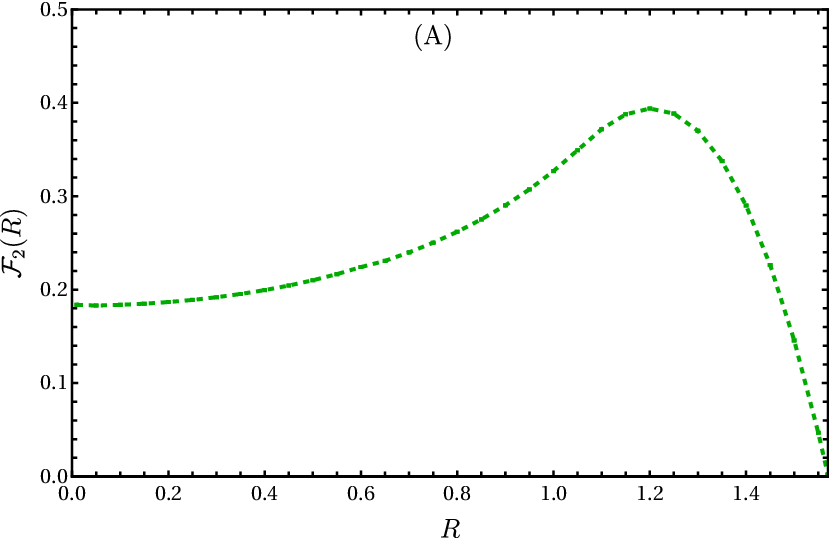}
\includegraphics[scale=0.53]{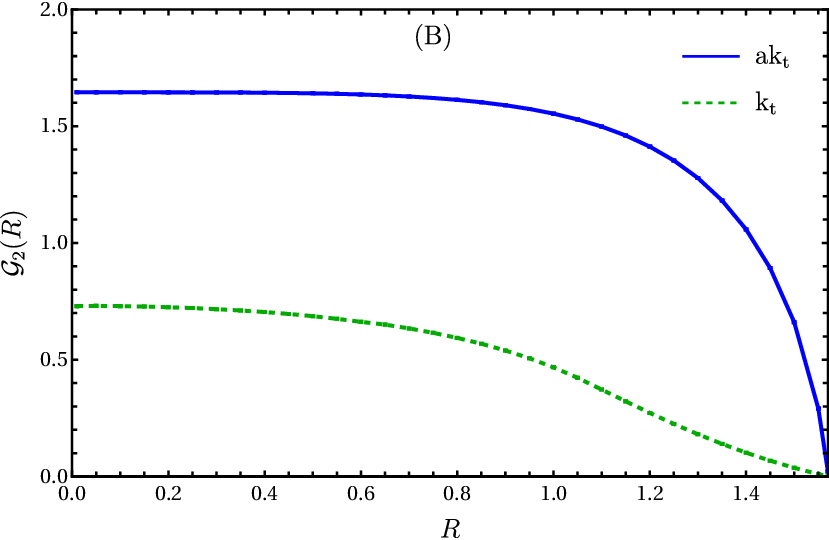}
\caption{The two-loop CLs and NGLs coefficients for the $k_t$ jet algorithm (from Ref.~\cite{Khelifa-Kerfa:2024hwx}).}
\label{fig:KT:F2-G2}
\end{figure}

The one- and two-loop results are given by \cite{Khelifa-Kerfa:2011quw, Khelifa-Kerfa:2024hwx}:
\begin{subequations}
\begin{align}\label{eq:1loop}
 \S_1(\rho) &= -\CF \asb \left[ L^2 - \left(\frac{3}{2} - 2 \ln \frac{1-c_R}{1+c_R}  \right) L \right],
\end{align}
where $L = \ln(1/\rho)$, and
\begin{align}
 \S_2(\rho) &= \frac{1}{2!} \left(\S_1\right)^2 + \S_{2,\cl}^{\ktt,\ca} + \S_{2,\ng}^{\ktt,\ca},
\end{align}
with the CLs and NGLs distributions given by:
\begin{align}
 \S_{2,\cl}^{\ktt,\ca}(\rho) &= 2\asb^2 \frac{L^2}{2!}\,\CFsq\, \F_2(R),
\label{eq:2loop:CLs} \\
 \S_{2,\ng}^{\ktt,\ca}(\rho) &=-2\asb^2 \frac{L^2}{2!}\,\CF\CA\, \G_2(R),
\label{eq:2loop:NGLs}
\end{align}
where $\F_2(R)$ and $\G_2(R)$ are the two-loop CLs and NGLs coefficients, respectively, plotted in Fig.~\ref{fig:KT:F2-G2}. The colour factors are $\CF = (N_c^2-1)/2N_c$ and $\CA = N_c$, where $N_c$ denotes the number of colours.
\end{subequations}
Recall that correlated emissions from gluons inside the quark jet that end up in the antiquark jet, or vice versa, yield non-global logarithms that are subleading, as shown in Refs.~\cite{Khelifa-Kerfa:2011quw, Kerfa:2012yae}.

\subsection{Three loops}
\label{sec:3loop}

This is the first order at which the C/A and $k_t$ jet algorithms begin to differ. For the emission of three soft, strong energy-ordered gluons, there are six distance measures to compare: $d_{1q}, d_{2q}, d_{3q}, d_{12}, d_{13}$, and $d_{23}$. We consider only one jet (the quark jet), though a similar treatment applies to the antiquark jet. Table~\ref{tab:3loop:CAClustering} details all configurations before and after applying the C/A algorithm, where $\Delta_{ijq} \equiv \Theta\left(d_{jq} - d_{ij}\right)$ with $i<j$, and the letter C indicates a mis-cancellation between real emissions and their corresponding virtual corrections. Such mis-cancellations are the source of large logarithmic terms—either non-global logarithms, clustering logarithms, or both.

\begin{table}[]
\centering
\begin{tabular}{@{}c|c|c|c|c|c|c|c|c|c|c@{}}
\toprule
$\Delta_{12q}$ & $\Delta_{13q}$  & $\Delta_{23q}$ & Orderings & $1_{\inn}$ & $2_{\inn}$ & $3_{\inn}$  &  $(1, 2)_{\inn}$ & $(1, 3)_{\inn}$ & $(2, 3)_{\inn}$  & $(1, 2, 3)_{\inn}$\\
\midrule
 0 & 0 & 0 & - & 0 & 0 & C & 0 & C & C & C \\
 1 & 0 & 0 & - & 0 & 0 & C & 0 & C & C & C \\
 0 & 1 & 0 & - & 0 & 0 & C & 0 & C & C & C \\
 0 & 0 & 1 & - & 0 & 0 & C & 0 & C & C & C \\
 1 & 1 & 0 & - & 0 & 0 & C & 0 & C & C & C \\
 1 & 0 & 1 & $d_{12} < d_{23}$ & 0 & 0 & C & 0 & C & C & C \\
 1 & 0 & 1 & $d_{23} < d_{12}$ & 0 & 0 & C & 0 & C & C & C \\
 0 & 1 & 1 & - & 0 & 0 & C & 0 & C & C & C \\
 1 & 1 & 1 & - & 0 & 0 & C & 0 & C & C & C \\
\bottomrule
\end{tabular}
\caption{Clustering patterns for the C/A algorithm at three loops.} \label{tab:3loop:CAClustering}
\end{table}

A gluon $k_i$ is considered inside the measured jet (denoted $i_{\inn}$ in Table~\ref{tab:3loop:CAClustering}) if it satisfies the clustering condition~\eqref{eq:Def:ClusCond}; otherwise, it is outside the jet and not explicitly shown. For example, in the fifth column, $1_{\inn}$ indicates that only gluon $k_1$ is initially inside the jet (prior to C/A clustering), while $k_2$ and $k_3$ are outside. Similarly, in the penultimate column, only $k_2$ and $k_3$ are initially inside, whilst $k_1$ is outside. The remaining columns follow the same labelling convention.

Note that $\Delta_{ijq} = 1$ implies $d_{jq} > d_{ij}$, meaning gluons $k_i$ and $k_j$ are clustered first by the C/A algorithm. Since $k_j$ is softer than $k_i$, it is pulled toward the harder gluon's position—either into or out of the measured jet, depending on $k_i$'s initial location. This ``dragging" mechanism spoils the complete cancellation between real and virtual contributions to the jet mass observable, thereby generating large logarithms. Conversely, $\Delta_{ijq} = 0$ implies $d_{jq} < d_{ij}$, so gluon $k_j$ remains part of the measured (quark) jet.

The ``Orderings" column lists possible hierarchies of the distance measures $d_{ij}$ for which the corresponding $\Delta_{ijq} = 1$. For instance, in rows 6 and 7, $\Delta_{12q} = \Delta_{23q} = 1$ indicates that the two smallest distances are $d_{12}$ and $d_{23}$. As the C/A algorithm requires identifying the absolute minimum, we consider both orderings $d_{12} < d_{23}$ and $d_{23} < d_{12}$ in analytical calculations. When different orderings yield distinct outcomes, both are explicitly shown; a dash (---) indicates that the ordering does not affect the final clustering result, as in the last row where all $\Delta_{ijq} = 1$.

As an illustrative example, consider the scenarios in rows 6 and 7. Table~\ref{tab:3loop:CAClustering-Row6} shows the outcome of the C/A algorithm for all gluonic configurations at three loops, specifically for the ordering $d_{12} < d_{23}$. Here, ``RRR" denotes all three gluons as real, ``RRV" indicates only gluon~3 is virtual, and so forth. We reiterate that a gluon contributes to the measured jet mass only if it is real and clustered into the jet by the algorithm.
\begin{table}[t]
\centering
\begin{tabular}{@{}c|c|c|c|c|c|c|c|c|c@{}}
\toprule
 & RRR  & RRV & RVR & VRR & RVV  & VRV & VVR & VVV & sum \\
\midrule
$1_\inn$ & $(1,2)_\inn$ & $(1,2)_\inn$ & $(1)_\inn$ & none & $(1)_\inn$ & none & none & none & 0 \\
$2_\inn$ & none & none & none & $(2,3)_\inn$ & none & $(2)_\inn$ & none &  none & 0
\\
$3_\inn$ & $(3)_\inn$ & none & $(3)_\inn$ & none & none & none & $(3)_\inn$ &  none & C
\\
$(1, 2)_\inn$ & $(1,2)_\inn$  & $(1,2)_\inn$ & $(1)_\inn$ & $(2,3)_\inn$ & $(1)_\inn$ & $(2)_\inn$ & none &  none & 0
\\
$(1, 3)_\inn$ & $(1,2,3)_\inn$  & $(1,2)_\inn$ & $(1,3)_\inn$ & none & $(1)_\inn$ & none & $(3)_\inn$ &  none & C
\\
$(2, 3)_\inn$ & $(3)_\inn$  & none & $(3)_\inn$ & $(2,3)_\inn$ & none & $(2)_\inn$ & $(3)_\inn$ &  none & C
\\
$(1, 2, 3)_\inn$ & $(1,2,3)_\inn$  & $(1,2)_\inn$ & $(1,3)_\inn$ & $(2,3)_\inn$ & $(1)_\inn$ & $(2)_\inn$ & $(3)_\inn$ &  none & C
\\
\bottomrule
\end{tabular}
\caption{Outcome of C/A algorithm for various gluon configurations at three loops, for the specific case $d_{12} < d_{23}$.} \label{tab:3loop:CAClustering-Row6}
\end{table}

For the scenario where only gluon~$1$ is initially inside the measured jet ($1_\inn$) and all gluons are real, the condition $d_{12} < d_{23}$ combined with strong-energy ordering means that gluon~$1$ pulls gluon~$2$ into the jet. They form a subjet with four-momentum approximately equal to that of the harder gluon~$1$. In the second clustering step, only gluons~$1$ and~$3$ remain. Since $\Delta_{13q} = 0$ (row~6 of Table~\ref{tab:3loop:CAClustering}), these gluons cannot cluster, and the algorithm terminates. The final outcome is gluons~1 and~2 inside the measured jet and gluon~3 outside. The same outcome occurs for the RRV configuration, but since it involves one virtual gluon (gluon~3), its emission amplitude equals minus that of RRR in the eikonal limit. Consequently, the RRR and RRV configurations cancel exactly.
Following the same procedure, we find that although the RVR and RVV configurations contribute to the jet mass, they also cancel. The remaining configurations yield no contribution, resulting in a total sum of zero.

The $2_\inn$ case follows a similar analysis. We note that the VRR and VRV configurations cancel in the strong-energy ordering regime.

For the $3_\inn$ case in the RRR configuration, gluons~1 and~2 are outside the measured jet and merge into a subjet aligned with the four-momentum of the harder gluon~1. Since $\Delta_{13q} = 0$, gluon~3 is not pulled out of the jet by gluon~1 and thus contributes to the jet mass. This contribution cancels against the RVR configuration (where gluon~2 is virtual). However, the VVR configuration yields a similar contribution to RRR but lacks a counterterm to cancel against. This mismatch between real and virtual diagrams results in a non-zero total contribution, generating both NGLs and CLs.

For the $(1,2)_\inn$ case, one may verify that the sum is zero. Note that in the VRR configuration, since gluon~1 is virtual, the smallest distance is $d_{23}$, so gluon~3 is pulled into the jet by the harder gluon~2.

For the $(1,3)_\inn$ case, the RRR and RRV configurations cancel due to strong-energy ordering, as do RVR and RVV. In the VRR configuration, gluon~1 is virtual and unaffected by the algorithm, so the smallest distance becomes $d_{23}$. The harder gluon~2 then pulls the softer gluon~3 out of the jet, yielding no contribution. Consequently, the VVR configuration has no counterpart to cancel against and contributes to the jet mass, resulting in a non-zero total.

For the $(2,3)_\inn$ case, the VVR configuration lacks a counterterm and contributes to the jet mass. Finally, when all three gluons are inside the jet $(1,2,3)_\inn$, all configurations cancel except VVR, which causes a real-virtual mismatch and yields a non-zero total.

Table~\ref{tab:3loop:CAClustering-Row7} shows the outcomes for the ordering $d_{23} < d_{12}$ (row~7). This ordering occurs in the $k_t$ jet algorithm, where the smallest distance is between the two softest gluons (2 and~3). Although this scenario is already included in $k_t$ calculations \cite{Khelifa-Kerfa:2024hwx}, we analyse Table~\ref{tab:3loop:CAClustering-Row7} in detail as it was not explicitly studied in that reference.
\begin{table}[t]
\centering
\begin{tabular}{@{}c|c|c|c|c|c|c|c|c|c@{}}
\toprule
& RRR  & RRV & RVR & VRR & RVV  & VRV & VVR & VVV & sum \\
\midrule
 $1_\inn$ & $(1,2,3)_\inn$ & $(1,2)_\inn$ & $(1)_\inn$ & none & $(1)_\inn$ & none & none & none & 0
\\
 $2_\inn$ & none & none & none & $(2,3)_\inn$ & none & $(2)_\inn$ & none & none & 0
\\
 $3_\inn$ & none & none & $(3)_\inn$ & none & none & none & $(3)_\inn$ & none & C
\\
 $(1,2)_\inn$ & $(1,2,3)_\inn$ & $(1,2)_\inn$ & $(1)_\inn$ & $(2,3)_\inn$ & $(1)_\inn$ & $(2)_\inn$ & none & none & 0
\\
 $(1,3)_\inn$ & $(1,2,3)_\inn$ & $(1,2)_\inn$ & $(1)_\inn$ & none & $(1)_\inn$ & none & $(3)_\inn$ & none &  C
\\
 $(2,3)_\inn$ & none & none & $(3)_\inn$ & $(2,3)_\inn$ & none & $(2)_\inn$ & $(3)_\inn$ & none  & C
\\
 $(1,2,3)_\inn$ & $(1,2,3)_\inn$ & $(1,2)_\inn$ & $(1,3)_\inn$ & $(2,3)_\inn$ & $(1)_\inn$ & $(2)_\inn$ & $(3)_\inn$ & none  &  C
\\
\bottomrule
\end{tabular}
\caption{Outcome of C/A algorithm for various gluon configurations at three loops, for the specific case $d_{23} < d_{12}$.} \label{tab:3loop:CAClustering-Row7}
\end{table}

For the case where only gluon~1 is inside the jet $(1_\inn)$ in the RRR configuration, gluons~2 and~3 cluster first, forming a subjet aligned with gluon~2's four-momentum. This subjet then clusters with gluon~1, bringing all three gluons into the jet. This contribution cancels against RRV due to strong-energy ordering. In the RVR configuration, gluons~1 and~3 cannot cluster ($\Delta_{13q} = 0$), so only gluon~1 remains in the jet. The same outcome occurs for RVV but with opposite sign, leading to cancellation. The total sum is zero. The $(2_\inn)$ case is trivial and also gives zero.

For the $(3_\inn)$ case, only the RVR and VVR configurations contribute to the jet mass. Their eikonal amplitudes differ, so they do not cancel in general, particularly for NGLs. However, for CLs (from the primary emission part of the eikonal amplitudes), they do cancel \cite{Delenda:2012mm, Delenda:2015tbo, Khelifa-Kerfa:2024hwx, Khelifa-Kerfa:2024dut}. Since our formalism includes both NGLs and CLs, we retain this as a real-virtual mismatch, resulting in a non-zero total.

The $(1,2)_\inn$ case leads to complete cancellation among real and virtual diagrams, yielding no contribution. For $(1,3)_\inn$ in the RRR configuration, gluon~2 pulls the softest gluon~3 out of the jet during the first clustering. The resulting subjet (aligned with gluon~2) is then pulled into the jet by gluon~1. This contribution cancels against RRV. The RVR and RVV contributions cancel, as do VRR and VRV, leaving only VVR uncancelled and yielding a non-zero total.

For the $(2,3)_\inn$ case, VRR and VRV cancel under strong-energy ordering, but RVR and VVR do not due to different eikonal amplitudes, creating a real-virtual mismatch that contributes to the jet mass. Finally, for $(1,2,3)_\inn$, all configurations cancel except VVR, which contributes to the jet mass.

The following points are worth noting:
\begin{itemize}
\item Cases leading to real-virtual mismatches (and thus large logarithms) are those where the \textit{softest} gluon is initially inside the jet region.
\item The mis-cancellation can arise from a single diagram or multiple diagrams with different eikonal emission amplitudes.
\item The difference between C/A and $k_t$ clustering arises when the two harder gluons cluster first, while the hardest and softest gluons cannot cluster. This prevents the softest gluon from being dragged in or out, unlike in $k_t$ where the softest gluon is always affected.
\end{itemize}

Summing over all gluon configurations in Table~\ref{tab:3loop:CAClustering} for the emission of three soft, energy-ordered gluons, we find that
\begin{align}\label{eq:3loop:uWX}
 \sum_\X \Uh_3 \, \W_{123}^{\X, \ca} &= \sum_\X \Uh_3 \, \W_{123}^{\X, k_t} - \Theta_1^\out \Theta_3^\inn \Theta_3^\rho \O_{12} \O_{23} \Ob_{13} \Delta_{123} \left[\W_{123}^{\V\V\R} + \W_{123}^{\R\V\R} + \W_{123}^{\R\R\R} \right],
\end{align}
where the first term corresponds to the $k_t$ jet clustering result, already computed in Ref.~\cite{Khelifa-Kerfa:2024hwx} (eq.~(52)). The constraint $\O_{ij}$ is defined as the Heaviside function $\Theta( d_{jB} - d_{ij})$. When $\O_{ij} = 1$, the parton pair $(ij)$ is closer to each other than parton $j$ is to the beam direction.

As explained earlier, the effect of the C/A algorithm on the jet mass distribution incorporates that of the $k_t$ algorithm, since it involves all possible orderings—only a subset of which are covered by the $k_t$ algorithm. Substituting into eq.~\eqref{eq:Def:Xsec-mOrder} yields:
\begin{align}\label{eq:3loop:Sigma}
\S_3^{\ca}(\rho) &= \S_3^{k_t}(\rho) + \widetilde{\S}_3^{\ca}(\rho), \notag\\
				 &= \frac{1}{3} \left(\S_1\right)^3 + \S_1 \times \left(\S_{2, \cl}^{\ca} + \S_{2, \ng}^{\ca}\right) + \S_{3, \cl}^{\ca} + \S_{3, \ng}^{\ca},
\end{align}
where $\S_1$, $\S_{2,\cl}^{\ca}$, and $\S_{2, \ng}^{\ca}$ denote the one-loop~\eqref{eq:1loop}, two-loop CLs~\eqref{eq:2loop:CLs}, and two-loop NGLs~\eqref{eq:2loop:NGLs} contributions to the jet mass fraction, respectively. The new three-loop CLs and NGLs contributions are given by the final two terms:
\begin{align}
 \S^{\ca}_{3, \cl}(\rho) &=  \S^{\ktt}_{3, \cl}(\rho) + \widetilde{\S}^{\ca}_{3, \cl}(\rho)
 = -2 \asb^3 \, \frac{L^3}{3!}\, \CFcub\, \left[\F^{\ktt}_3 + \widetilde{\F}_3^{\ca} \right],
\label{eq:3loop:CLs}
\\
 \S^{\ca}_{3, \ng}(\rho) &=  \S^{\ktt}_{3, \ng}(\rho) + \widetilde{\S}^{\ca}_{3, \ng}(\rho)
 = +2 \asb^3 \, \frac{L^3}{3!}\, \left[\CFsq \CA \left(\G^{\ktt}_{3,a} + \widetilde{\G}_{3,a}^{\ca}\right) + \CF \CAsq \left(\G^{\ktt}_{3,b} + \widetilde{\G}_{3,b}^{\ca}\right)\right],
\label{eq:3loop:NGLs}
\end{align}
The $k_t$ results for both CLs and NGLs have been presented in Ref.~\cite{Khelifa-Kerfa:2024hwx} (eqs.~(55), (56), and (57), and Fig.~4). The new C/A clustering corrections are given by:
\begin{align}
 \widetilde{\F}_{3}^{\ca}(R) &= \int_{1_\out} \int_{3_\inn} \O_{12} \O_{23} \Ob_{13}\, \Theta\left(d_{23} - d_{12}\right) \, w_{ab}^1 w_{ab}^2 w_{ab}^3,
\end{align}
for the CLs coefficient, and
\begin{subequations}\label{eq:3loop:G3}
\begin{align}
 \widetilde{\G}_{3,a}^{\ca}(R) &= -\frac{1}{2} \int_{1_\out} \int_{3_\inn} \O_{12} \O_{23} \Ob_{13} \Theta\left(d_{23} - d_{12}\right) \, w_{ab}^1\, \A_{ab}^{23},
\label{eq:3loop:G3a}
\\
 \widetilde{\G}_{3,b}^{\ca}(R) &= -\frac{1}{4} \int_{1_\out} \int_{3_\inn} \O_{12} \O_{23} \Ob_{13} \Theta\left(d_{23} - d_{12}\right) \, \B_{ab}^{123},
\label{eq:3loop:G3b}
\end{align}
\end{subequations}
for the NGLs coefficients.

\begin{figure}
\centering
\includegraphics[scale=0.65]{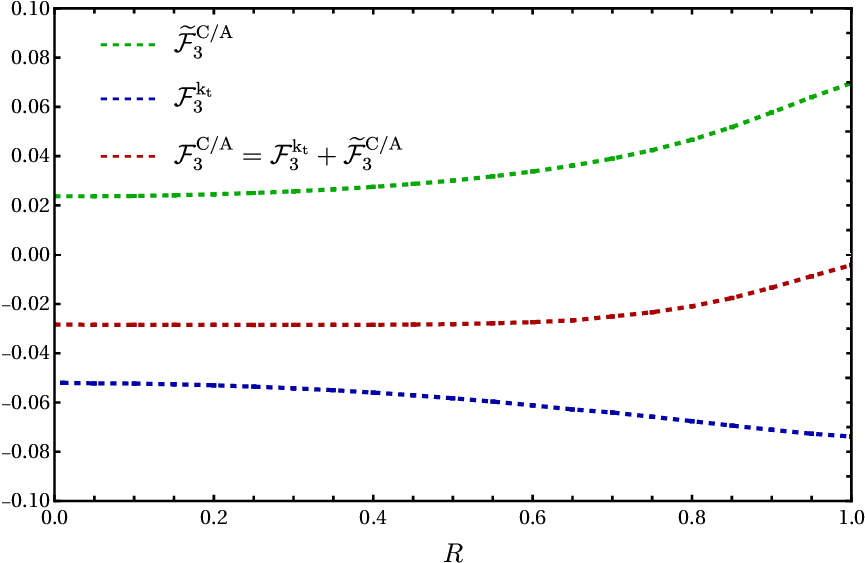}
\caption{The three-loop CLs coefficient for the C/A and $k_t$ jet algorithms.}
\label{fig:CA:F3}
\end{figure}
In Fig.~\ref{fig:CA:F3}, we plot $\F_3^{k_t}$, $\widetilde{\F}_3^{\ca}$, and their sum $\F_3^{\ca} = \widetilde{\F}_3^{\ca} + \F_3^{k_t}$, which represents the CLs coefficient for the C/A algorithm at three loops. In the limit $R \to 0$, we obtain $\F_3^{\ca}(0) = -0.028$, confirming our previous calculations \cite{Delenda:2012mm}. The persistence of CLs even as the jet radius shrinks to zero is a phenomenon known as the \textit{edge} or \textit{boundary} effect, previously observed for both CLs and NGLs in anti-$k_t$ and $k_t$ algorithms (see, e.g., Refs.~\cite{Dasgupta:2002bw, Kerfa:2012yae, Khelifa-Kerfa:2011quw, Khelifa-Kerfa:2024hwx}). Fig.~\ref{fig:CA:F3} clearly shows that this behaviour also occurs for the C/A algorithm. Moreover, the small-$R$ limit of the CLs coefficient for our jet mass observable is both much smaller and of opposite sign compared to that reported for the hemisphere mass observable \cite{Khelifa-Kerfa:2025cdn}, which is $+1.363$. An explanation for this difference is provided in the latter reference. Similar observations were reported for the two-loop results for both CLs and NGLs.

C/A clustering reduces the $k_t$ CLs coefficients by $45\%$ at $R=0$ to approximately $95\%$ at $R=1$. Additionally, unlike $k_t$, the magnitude of the C/A CLs coefficients decreases with increasing $R$ (red curve). Since CLs are notoriously difficult to compute even at the first few orders of perturbative expansion and at leading logarithmic accuracy, minimising their impact—or removing them entirely—would be highly desirable. These two features make the C/A algorithm preferable to $k_t$ (at least at this loop order). Indeed, at $R=1$:
\begin{align}
 \S_{3,\cl}^{\ca}(\rho) = +\asb^3 \,L^3\,  \left(0.0033\right), \qquad
 \S_{3,\cl}^{\ktt}(\rho) = +\asb^3 \,L^3\, \left(0.058\right).
\end{align}
Comparing this to the leading term $\asb^3\,L^6\, (0.39)$, the CLs contribution is negligible, accounting for less than $0.8\%$ of the leading term (compared to approximately $15\%$ for $k_t$ CLs).

\begin{figure}
\centering
\includegraphics[scale=0.53]{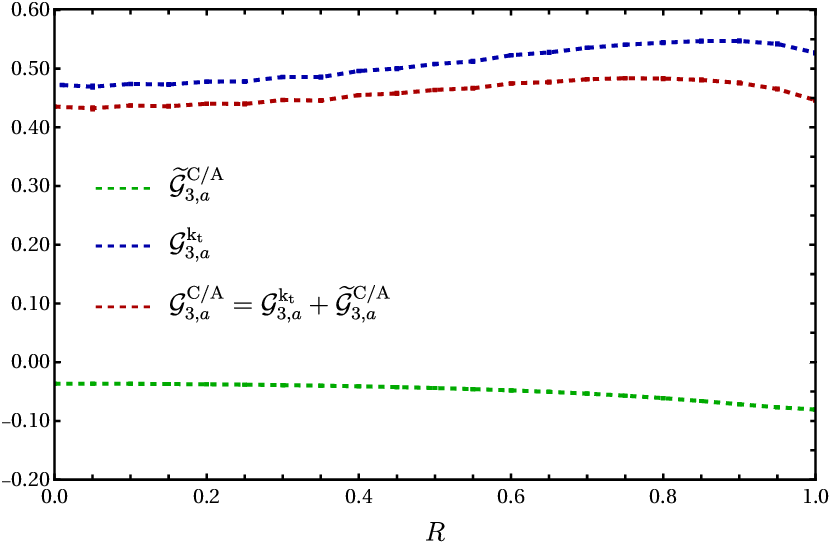}
\includegraphics[scale=0.53]{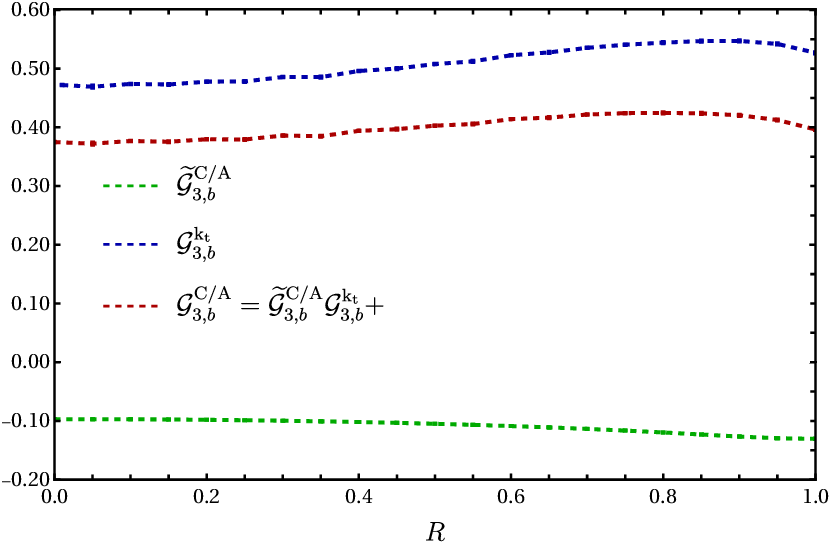}
\caption{The three-loop NGLs coefficients for the C/A and $k_t$ jet algorithms.}
\label{fig:CA:G3ab}
\end{figure}

In Fig.~\ref{fig:CA:G3ab}, we plot the results of the integrals in Eqs.~\eqref{eq:3loop:G3} alongside the corresponding $k_t$ results from Ref.~\cite{Khelifa-Kerfa:2024hwx}. The sums $\G_{3,i}^{\ca} = \widetilde{\G}_{3,i}^{\ca} + \G_{3,i}^{k_t}$ (where $i=a,b$), representing the total C/A NGLs coefficients at this loop order, are also plotted.

The C/A corrections to the $k_t$ results are relatively small. However, similar to the CLs case, these corrections have the opposite sign to the $k_t$ coefficients, thus reducing the final C/A NGLs coefficients. The reduction ranges from approximately $8\%$ at $R=0$ to $18\%$ at $R=1$ for the $\CFsq \CA$ coefficient ($\G_{3,a}$), and from $26\%$ at $R=0$ to $33\%$ at $R=1$ for the $\CF \CAsq$ coefficient ($\G_{3,b}$). Similar to CLs, the magnitude of NGLs is smaller in C/A than in $k_t$ clustering. Given the extreme complexity of calculating both CLs and NGLs, this makes C/A preferable to $k_t$, as an ideal jet algorithm should minimise the impact of non-global effects (whether CLs or NGLs).

In the small-$R$ limit, we obtain $\G_{3,a}^{\ca}(0) = +0.37$ and $\G_{3,b}^{\ca}(0) = +0.57$. These values differ from their hemisphere mass counterparts \cite{Khelifa-Kerfa:2025cdn}: $\G_{3,a}^{\ca, \mathrm{hemi}} = +0.655$ and $\G_{3,b}^{\ca, \mathrm{hemi}} = +0.198$. This discrepancy may be due to differences in the geometric distortion of the jet boundary across algorithms, as explained in the latter reference. Since CLs and NGLs manifestly originate from the jet boundary, its geometric shape plays a crucial role in determining their magnitude.

\begin{figure}
\centering
\includegraphics[scale=0.65]{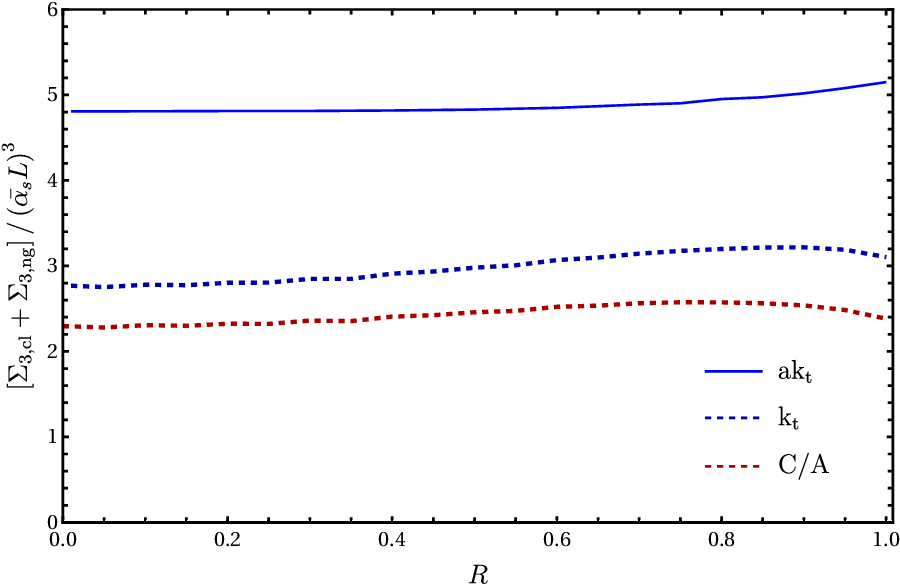}
\caption{The sum of CLs and NGLs contributions to the jet mass fraction at three loops for three jet algorithms.}
\label{fig:CA:G3F3}
\end{figure}

To assess the impact of the C/A and $k_t$ algorithms on the jet mass distribution, we compare them to the anti-$k_t$ algorithm, which is free of CLs but exhibits relatively large NGLs. This comparison is shown in Fig.~\ref{fig:CA:G3F3}, where we sum both CLs and NGLs contributions (Eqs.~\eqref{eq:3loop:CLs} and \eqref{eq:3loop:NGLs}) including colour factors. The anti-$k_t$ and $k_t$ results are taken from our previous paper \cite{Khelifa-Kerfa:2024hwx}.
The reduction in the anti-$k_t$ NGLs coefficient varies slightly for both $k_t$ and C/A over the jet radii considered. For $k_t$, it ranges from $42\%$ ($R=0$) to $40\%$ ($R=1$), while for C/A it increases from $52\%$ ($R=0$) to $54\%$ ($R=1$). The C/A clustering algorithm reduces the total non-global contribution to the jet mass distribution by more than $50\%$ compared to anti-$k_t$, making it preferable to both anti-$k_t$ and $k_t$ for mitigating non-global effects at three loops.

In the next section, we present calculations of the four-loop CLs and NGLs coefficients for the jet mass observable. Together with the three-loop NGLs, these represent the state-of-the-art in fixed-order C/A clustering calculations and, to the best of our knowledge, have not been previously reported in the literature.

\subsection{Four loops}
\label{sec:4loop}

For the emission of four soft, energy-ordered gluons, applying the C/A algorithm becomes considerably more complex, yielding lengthy and intricate expressions. To manage this complexity, we automated the entire process using a \texttt{Python} code. This code generates a comprehensive table containing all possible configurations, their corresponding orderings, and the resulting expressions for the sum of eikonal amplitudes that contribute to the mis-cancellation between real emissions and virtual corrections. The code is readily generalisable to higher loop orders. In total, there are $2^6 = 64$ possible configurations with 1957 distinct orderings.

Table~\ref{tab:4loop:CAClustering} presents a sample of the code output (with explicit eikonal amplitude expressions replaced by 0 or C, as in the three-loop case). We show only the gluonic scenarios that exhibit mis-cancellations. The remaining scenarios not explicitly shown are: $(1,2,4)_\inn$, $(1,3,4)_\inn$, $(2,3,4)_\inn$, and $(1,2,3,4)_\inn$. Scenarios free of such mis-cancellations—and thus yielding no large logarithms—include cases such as $1_\inn$, $2_\inn$, $3_\inn$, $(1,2)_\inn$, and several others.

\begin{table}
\centering
\begin{tabular}{@{}c|c|c|c|c|c|c|c|c|c|c@{}}
\toprule
$\Delta_{12q}$ & $\Delta_{13q}$  & $\Delta_{14q}$ & $\Delta_{23q}$ & $\Delta_{24q}$ & $\Delta_{34q}$ & Orderings & $4_{\inn}$ & $(3,4)_{\inn}$ & $(2,4)_{\inn}$  &  $(1, 4)_{\inn}$ \\
\midrule
 0 & 0 & 0 & 0 & 1 & 1 & $d_{24} < d_{34}$ & C & C & C & C \\
 0 & 0 & 0 & 0 & 1 & 1 & $d_{34} < d_{24}$ & C & C & C & C
\\
 0 & 0 & 0 & 1 & 1 & 1 & $d_{23} < d_{24}<d_{34}$ & C & C & C & C \\
 0 & 0 & 0 & 1 & 1 & 1 & $d_{23} < d_{34}<d_{24}$ & C & C & C & C \\
 0 & 0 & 0 & 1 & 1 & 1 & $d_{24} < d_{23}<d_{34}$ & C & C & C & C \\
 0 & 0 & 0 & 1 & 1 & 1 & $d_{24} < d_{34}<d_{23}$ & C & C & C & C \\
 0 & 0 & 0 & 1 & 1 & 1 & $d_{34} < d_{23}<d_{24}$ & C & C & C & C \\
 0 & 0 & 0 & 1 & 1 & 1 & $d_{34} < d_{24}<d_{23}$ & C & C & C & C \\
\bottomrule
\end{tabular}
\caption{A sample of C/A algorithm results at four loops.} \label{tab:4loop:CAClustering}
\end{table}

To illustrate the C/A algorithm at four loops, consider the third row of Table~\ref{tab:4loop:CAClustering} for case $4_\inn$ and ordering $d_{23} < d_{24} < d_{34}$. The 16 gluon configurations (RRRR, RRRV, RRVR, RVRR, etc.) yield the following clustering outcomes:
\begin{enumerate}
\item \textbf{RRRR}: Gluons 2 and 3 merge first into a subjet aligned with gluon~2 (due to strong ordering). Gluon~4 is then pulled out of the jet and merged with gluon~2, forming a subjet again aligned with gluon~2. The final configuration (gluons~1 and~2 outside the jet) yields no jet mass contribution. The RRRV configuration has identical but opposite eikonal amplitude and cancels with RRRR.

\item \textbf{RRVR}: Gluon~3 is virtual and unaffected. Gluons~2 and~4 merge into a subjet aligned with gluon~2, leaving gluons~1 and~2 outside the jet—no mass contribution. This cancels with RRVV.

\item \textbf{RVRR}: Gluon~3 pulls gluon~4 out of the jet, forming a subjet aligned with gluon~3. Gluons~1 and~3 remain outside the jet—no mass contribution. This cancels with RVRV.

\item \textbf{VRRR}: Identical to RRRR, with no mass contribution. Cancels with VRRV.

\item \textbf{VVRR}: Gluon~3 pulls gluon~4 out of the jet—no mass contribution. Cancels with VVRV.

\item \textbf{VRVR}: Gluon~2 pulls gluon~4 out of the jet—no mass contribution. Cancels with VRVV.

\item \textbf{RVVR}: Gluons~2 and~3 are virtual and cannot affect gluon~4, which remains inside the jet and contributes to its mass. The counterpart RVVV has no contribution (gluon~4 virtual), resulting in a real–virtual mis-cancellation.

\item \textbf{VVVR}: Identical to RVVR, contributing to the jet mass. Not cancelled by VVVV (gluon~4 virtual).
\end{enumerate}
The final expression for this case (third row, $4_\inn$) is:
\begin{align}
 - \Theta_1^\out \Theta_2^\out \Theta_3^\out \Theta_4^\inn\, \Theta_4^\rho \, \Theta\left(d_{34} - d_{24}\right) \Theta\left(d_{24} - d_{23}\right)\, \left[\W_{1234}^{\R\V\V\R} + \W_{1234}^{\V\V\V\R} \right].
\end{align}
This calculation is repeated for all 1957 cases. For each configuration, we sum contributions from all orderings, yielding 64 total expressions. These tasks are performed by the \texttt{Python} code. Using \texttt{Mathematica} and the full $k_t$ contribution from \cite{Khelifa-Kerfa:2024hwx}, we subtract the $k_t$ component to isolate new C/A clustering contributions. We substitute explicit eikonal amplitude expressions from ref.~\cite{Delenda:2015tbo}, split coefficients by colour factor, and compute them numerically using the \texttt{Cuba} library \cite{Hahn:2004fe}.

\begin{figure}[t]
\centering
\includegraphics[scale=0.65]{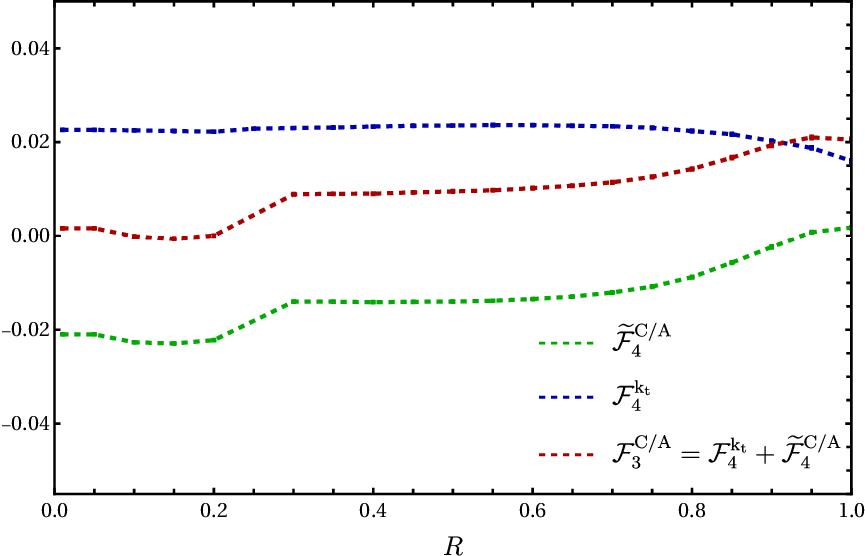}
\caption{The four-loop CLs coefficient for the C/A and $k_t$ jet algorithms.}
\label{fig:CA:F4}
\end{figure}
\begin{figure}[h]
\centering
\includegraphics[scale=0.53]{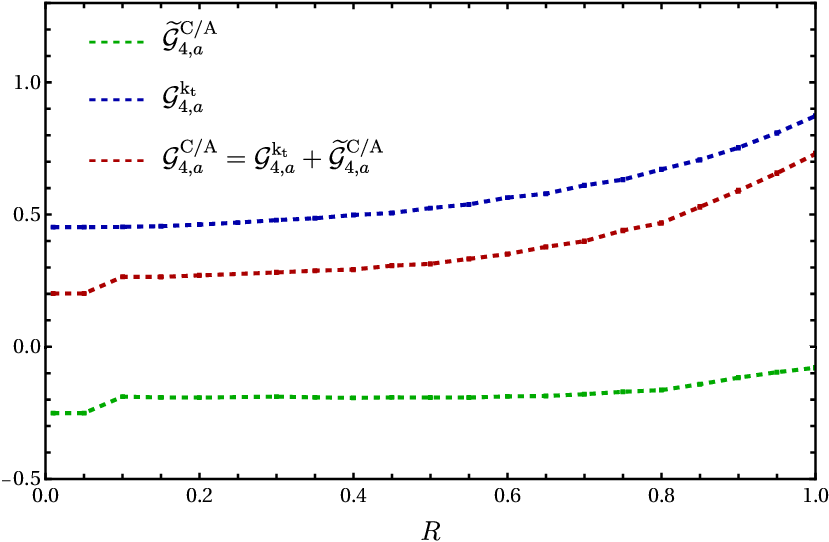}
\includegraphics[scale=0.53]{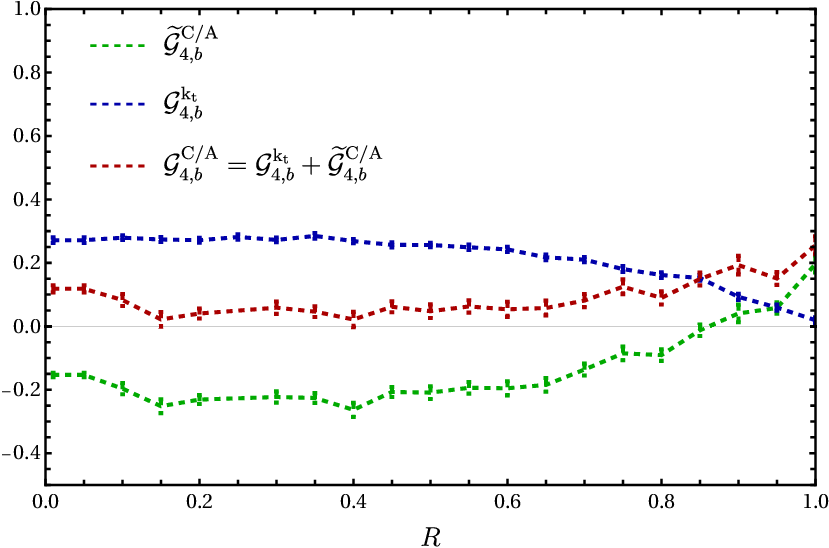}
\\
\includegraphics[scale=0.53]{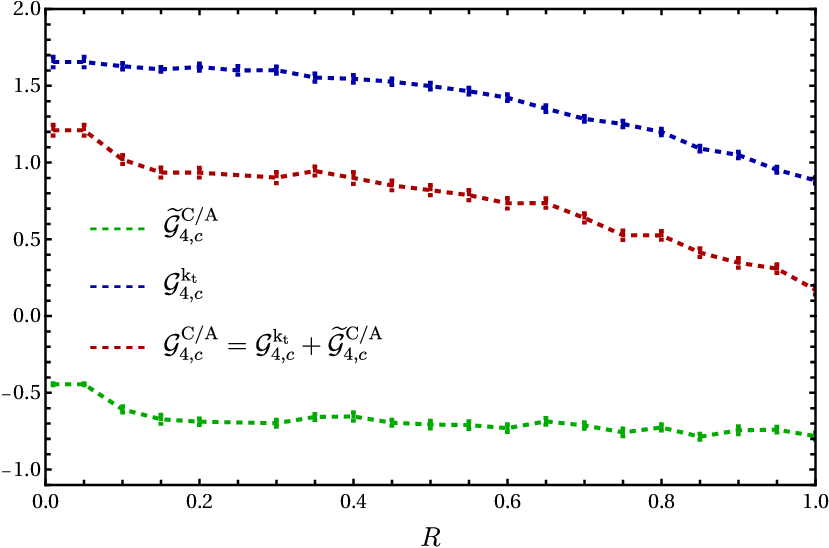}
\includegraphics[scale=0.53]{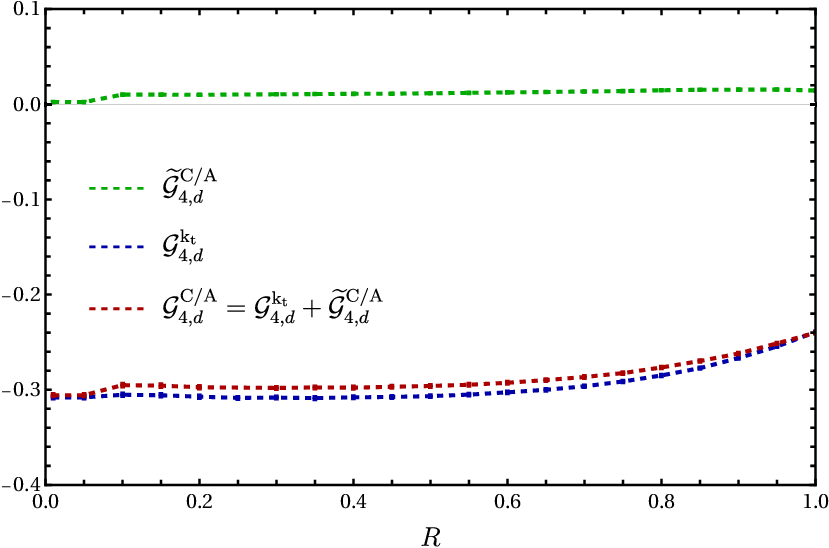}
\caption{The four-loop NGLs coefficients for the C/A and $k_t$ jet algorithms.}
\label{fig:CA:G4abcd}
\end{figure}

The jet mass fraction at four loops takes a form analogous to the three-loop case \eqref{eq:3loop:Sigma}:
\begin{align}\label{eq:4loop:Sigma}
 \S_4^{\ca}(\rho) &= \S_4^{\ktt} + \widetilde{\S}_4^{\ca},
\notag\\
&= \frac{1}{4!} \left(\S_1\right)^4 + \S_1 \times \left(\S_{3,\cl}^{\ca} + \S_{3,\ng}^{\ca} \right) + \frac{1}{2!} \left(\S_1\right)^2 \times \left(\S_{2,\cl}^{\ca} + \S_{2, \ng}^{\ca}\right) + \frac{1}{2!} \left(\S_{2,\cl}^{\ca}\right)^2
\notag\\
&+ \frac{1}{2!} \left(\S_{2,\ng}^{\ca}\right)^2 + \S_{4, \cl}^{\ca} + \S_{4, \ng}^{\ca},
\end{align}
where the new C/A clustering contributions are (see Refs.~\cite{Khelifa-Kerfa:2024hwx, Khelifa-Kerfa:2025cdn}):
\begin{align}\label{eq:4loop:CLs}
 \S_{4, \cl}^{\ca}(\rho) = +2 \asb^4\,\frac{L^4}{4!}\, \CFfour\, \left[\F_4^{\ktt} + \widetilde{\F}_4^{\ca} \right],
\end{align}
for the CLs contribution, and:
\begin{align}\label{eq:4loop:NGLs}
 \S_{4, \ng}^{\ca}(\rho) &= -2\asb^4\,\frac{L^4}{4!} \Big[
-\CFcub\CA \left(\G_{4,a}^{\ktt} + \widetilde{\G}_{4,a}^{\ca} \right)
-\CFsq\CAsq \left(\G_{4,b}^{\ktt} + \widetilde{\G}_{4,b}^{\ca} \right) +
\CF\CAcub \left(\G_{4,c}^{\ktt} + \widetilde{\G}_{4,c}^{\ca} \right) +
\notag\\
&\hspace{50pt} + \CF\CAsq \left(\CA -2\CF\right) \left(\G_{4,d}^{\ktt} + \widetilde{\G}_{4,d}^{\ca} \right)
\Big],
\end{align}
for the NGLs contribution. The subscripts $a,b,c,d$ appearing in $\G_4^{\ktt(\ca)}$ in the equation above label the coefficients multiplying the colour factors $\CFcub \CA, \CFsq\CAsq, \CF \CAcub$ and $\CF\CAsq (\CA-2\CF)$, respectively.

Integration results as functions of the jet radius $R$ are shown in Figs.~\ref{fig:CA:F4} and \ref{fig:CA:G4abcd}. Apart from minor irregularities at very small $R$ values (attributable to limited statistics\footnote{We employed approximately $10^{11}$ integrand evaluations for the \texttt{Vegas} routine in the \texttt{Cuba} library.}), C/A clustering induces substantial reductions in both CLs and NGLs coefficients across most $R$ values. However, for $R \gtrsim 0.85$, the C/A coefficients—particularly $\F_4^{\ca}$ and $\G_{4,b}^{\ca}$—become comparable to or exceed their $k_t$ counterparts.

\begin{figure}
\centering
\includegraphics[scale=0.65]{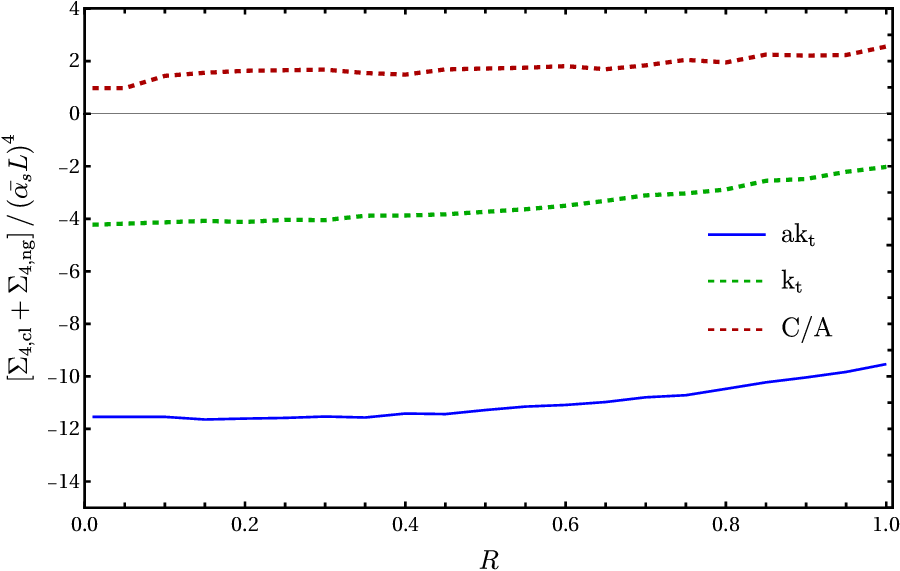}
\caption{Comparison of combined CLs and NGLs effects at four loops for three jet algorithms.}
\label{fig:CA:G4+F4}
\end{figure}

These observations are confirmed in Fig.~\ref{fig:CA:G4+F4}, which plots the combined CLs and NGLs contributions for C/A, $k_t$, and anti-$k_t$ algorithms. In addition to the sign change of the C/A contribution (positive, unlike the negative contributions of $k_t$ and anti-$k_t$), the reduction ranges from $92\%$ at $R=0$ to $74\%$ at $R=1$. For $k_t$, the reduction varies between $63\%$ and $78\%$ over the same range. At $R=1$, the $k_t$ reduction slightly exceeds that of C/A. Overall, we conclude that up to four loops at single-logarithmic accuracy, the C/A clustering algorithm should be prioritised over both anti-$k_t$ and $k_t$ for studies of non-global observables, as it minimises the impact of non-global logarithms (both CLs and NGLs).

Comparing with our previous hemisphere mass calculations \cite{Khelifa-Kerfa:2025cdn}, the small-$R$ limit values:
\begin{align}
 \F_4^{\ca} = 0.02, \quad \G_{4,a}^{\ca} = 0.2, \quad \G_{4,b}^{\ca} = 0.12, \quad \G_{4,c}^{\ca} = 1.21, \quad \G_{4,d}^{\ca} = -0.31,
\end{align}
differ significantly from those in eqs.~(49) and (52) of \cite{Khelifa-Kerfa:2025cdn}. This discrepancy may stem from geometric distortion of the jet boundary, as discussed previously.

In the next section, we discuss the exponentiation of CLs and NGLs for the C/A algorithm and compare their impacts on the full resummed jet mass distribution with those of anti-$k_t$ and $k_t$.

\section{All-order treatment}
\label{sec:all-order}

Since the C/A algorithm encompasses the $k_t$ algorithm, as explained in previous sections, it follows an identical structure and introduces only additional corrections absent in the latter. Therefore, the exponentiation of both CLs and NGLs reported in previous studies (see, for instance, \cite{Dasgupta:2001sh, Khelifa-Kerfa:2024roc, Khelifa-Kerfa:2024hwx, Khelifa-Kerfa:2025cdn, Khelifa-Kerfa:2025jev}) for anti-$k_t$ and $k_t$ algorithms also holds for C/A clustering. Accordingly, we write the full resummed jet mass distribution up to NLL accuracy in the standard form \cite{Dasgupta:2001sh, Kerfa:2012yae, Khelifa-Kerfa:2015mma, Khelifa-Kerfa:2024hwx, Khelifa-Kerfa:2025cdn}:
\begin{align}\label{eq:AllOrder:Full}
 \S^{\ca}(\rho) = \S^\P(\rho)\, \cS^{\ca}(\rho) \, \C^{\ca}(\rho),
\end{align}
where $\S^\P(\rho)$ is the familiar Sudakov form factor that resums leading double logarithms from soft and collinear emissions, given by \cite{Catani:1992ua}:
\begin{align}\label{eq:AllOrder:Primary}
 \S^\P(\rho) = \frac{\exp\left[-\cR - \gamma_E \cR'\right]}{\Gamma\left(1 + \cR'\right)},
\end{align}
where the radiator $\cR$ contains both leading and next-to-leading large logarithmic contributions to the jet mass fraction, $\cR'$ is its derivative with respect to $\tilde{L} = \ln(R^2/\rho)$, and $\gamma_E = 0.577$ is the Euler–Mascheroni constant. Note that at single-logarithmic accuracy, $\tilde{L} \sim L = \ln(1/\rho)$.
The full expressions for $\cR$ and its derivative can be found in, for instance, Refs.~\cite{Banfi:2010pa, Kerfa:2012yae, Khelifa-Kerfa:2025cdn}.

The NGLs form factor for C/A clustering, $\cS^{\ca}(\rho)$, resums the leading single-logarithmic NGLs to all orders. Unfortunately, there exists neither an analytical formula nor a numerical estimate of its size. The Monte Carlo code of Ref.~\cite{Dasgupta:2001sh} does not implement the C/A jet algorithm, though it can compute NGLs resummed form factor for soft and collinear observables such as the jet mass. Other numerical resummation approaches, such as \cite{Hatta:2013iba, Hagiwara:2015bia, Becher:2023znt}, either lack the C/A jet algorithm or are only valid for soft large-angle observables like gaps-between-jets.
We thus resort to estimating it using the fixed-order calculations up to four loops presented in previous sections, and the observed exponentiation pattern. Accordingly, we write, following analogous calculations for anti-$k_t$ and $k_t$ algorithms \cite{Delenda:2006nf, Khelifa-Kerfa:2015mma, Khelifa-Kerfa:2024hwx, Khelifa-Kerfa:2025cdn}:
\begin{align}\label{eq:AllOrder:NGLs}
 \cS^{\ca}(t) = \exp\left[ -2 \sum_{n \geq 2} \frac{1}{n!} \, \cS_n^{\ca}(R) \left(-2 \, t\right)^n \right],
\end{align}
where the $n$-loop NGLs coefficients are given, up to four loops, by:
\begin{align}\label{eq:AllOrder:NGLs-Coefs}
 &\cS_2^{\ca} = \CF \CA\, \G_2^{\ca}, \quad
 \cS_3^{\ca} = \CFsq \CA\, \G_{3,a}^{\ca} + \CF \CAsq\, \G_{3,b}^{\ca}, \notag\\
 &\cS_4^{\ca} = -\CFcub \CA\, \G_{4,a}^{\ca} - \CFsq \CAsq \,\G_{4,b}^{\ca} + \CF \CAcub \, \G_{4,c}^{\ca} + \CFsq \left(\CA - 2\CF\right) \,\G_{4,d}^{\ca}.
\end{align}
Here $t$ is the evolution variable, which at single-logarithmic accuracy is given by \cite{Dasgupta:2001sh, Banfi:2010pa, Khelifa-Kerfa:2024hwx}:
\begin{align}
 t = \frac{1}{2\pi} \int_{\rho/R^2}^1 \frac{\d x}{x} \, \as\left(x Q/2\right) = -\frac{1}{4\pi \beta_0} \, \ln\left[1 - 2 \as \beta_0 \tilde{L}\right],
\end{align}
where $\as \equiv \as(Q R/2)$ and the last equality shows the one-loop expansion of the integral, sufficient for our NLL accuracy. Note that for a fixed coupling, $t = \asb \tilde{L}/2$. Furthermore, expanding \eqref{eq:AllOrder:NGLs} yields the NGLs contributions at two, three, and four loops (Eqs.~\eqref{eq:2loop:NGLs}, \eqref{eq:3loop:NGLs}, and \eqref{eq:4loop:NGLs}, respectively).

The CLs form factor for the C/A algorithm, $\C^{\ca}(\rho)$, resums the leading single-logarithmic clustering logarithms for the jet mass observable. Similar to the NGLs form factor, there exists neither an analytical formula nor a numerical estimate for it. Since it also exhibits an exponentiation pattern, we can express it as an exponential of the fixed-order results. To this end, we write it in a form analogous to that for NGLs \eqref{eq:AllOrder:NGLs}:
\begin{align}\label{eq:AllOrder:CLs}
 \C^{\ca}(t) = \exp\left[ 2 \sum_{n \geq 2} \frac{1}{n!} \, \F_n^{\ca}(R) \left(-2\,\CF\, t \right)^n \right],
\end{align}
where the CLs coefficients at two, three, and four loops are given in Eqs.~\eqref{eq:2loop:CLs}, \eqref{eq:3loop:CLs}, and \eqref{eq:4loop:CLs}, respectively.

As stated above, there are no all-orders results for the jet mass observable—either numerical or analytical—in either the finite-$N_c$ or large-$N_c$ limits. To quantify the impact of the NGLs and CLs exponentials \eqref{eq:AllOrder:NGLs} and \eqref{eq:AllOrder:CLs}, we compare them to the output of the MC code \cite{Dasgupta:2001sh} for the anti-$k_t$ and $k_t$ algorithms. For this purpose, we use the well-known parametrisations (see, for instance, Refs.~\cite{Dasgupta:2001sh, Delenda:2006nf, Khelifa-Kerfa:2011quw, Kerfa:2012yae}):
\begin{align}\label{eq:AllOorder:NGLs-MC}
 \cS_{\MC}(t) &= \exp\left[-\CF \CA\, \G_2(R)\, \left(\frac{1 + (a t)^2}{1 + (b t)^c}\right) t^2\right],
\end{align}
for the NGLs form factor, and
\begin{align}\label{eq:AllOorder:CLs-MC}
 \C_{\MC}(\rho) &= \exp\left[\CFsq \,\F_2(R)\, \left(\frac{1 + (a t)^2}{1 + (b t)^c}\right) t^2\right],
\end{align}
for the CLs form factor. The values of the two-loop coefficients $\G_2(R)$ and $\F_2(R)$, along with the fitting parameters $a$, $b$, and $c$, are presented in Tables~1 and~2 of Ref.~\cite{Khelifa-Kerfa:2024hwx} for $R = 0.7$ and $1.0$.

\begin{figure}
\centering
\includegraphics[scale=0.51]{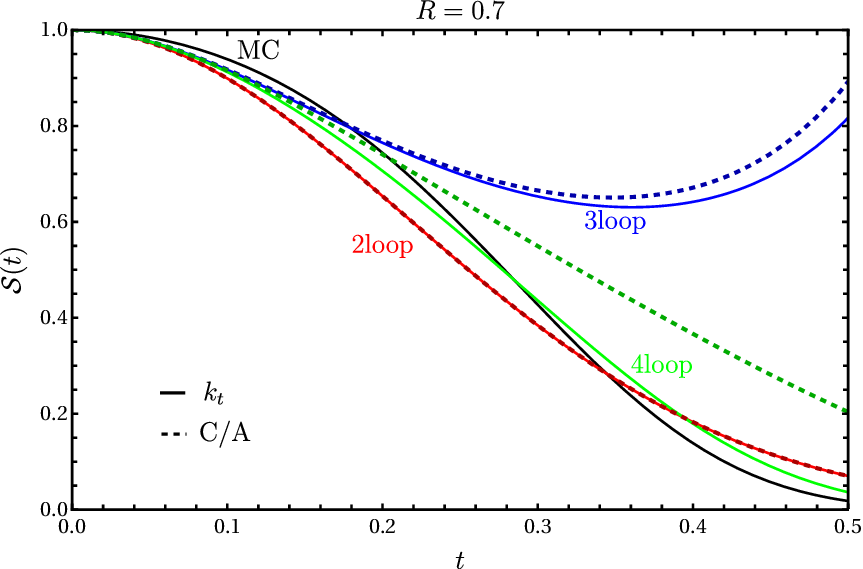}
\includegraphics[scale=0.51]{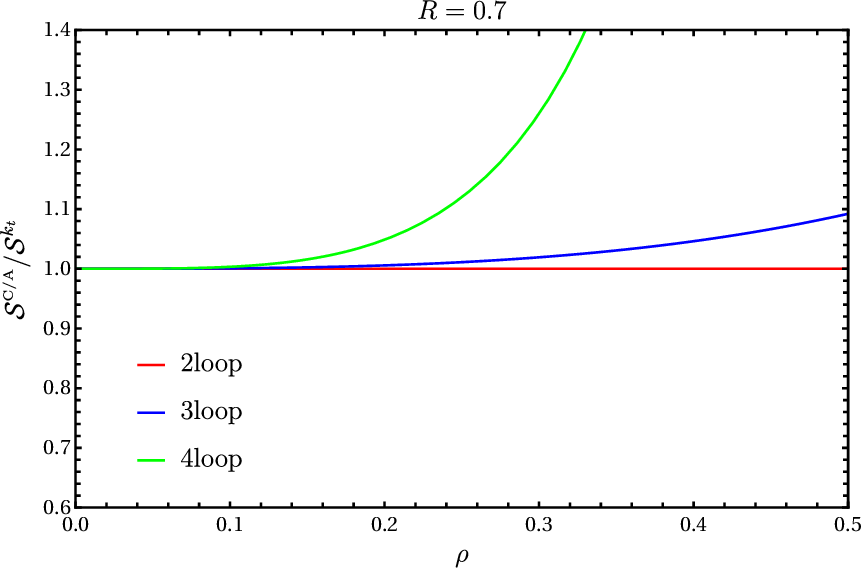}
\vskip 5pt
\includegraphics[scale=0.51]{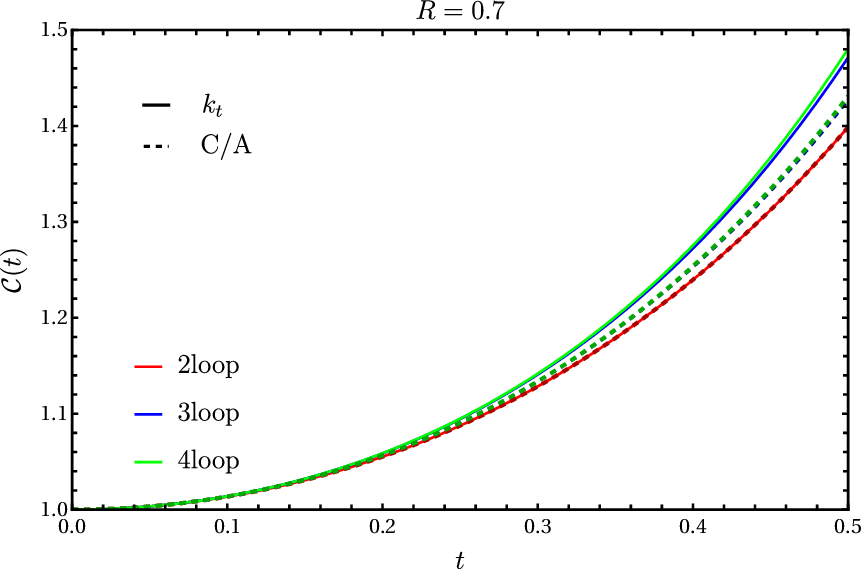}
\includegraphics[scale=0.51]{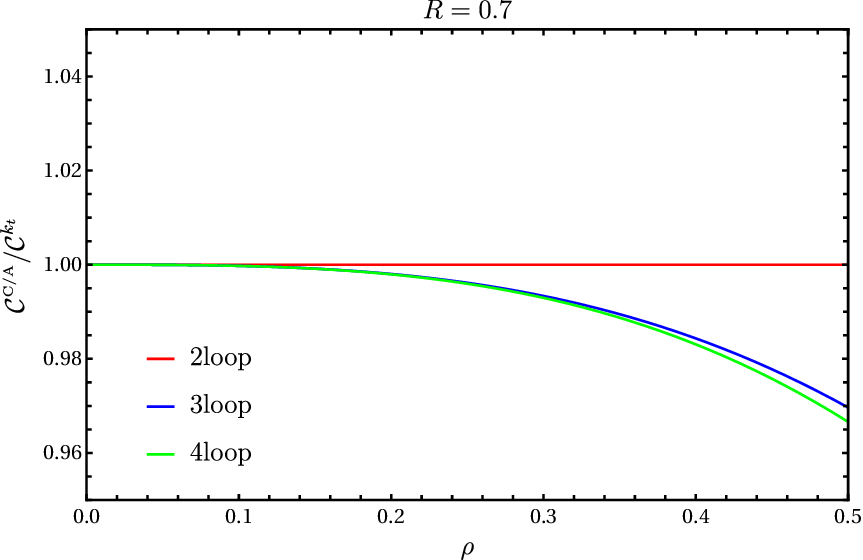}
\caption{Comparison of the exponentials $\cS(t)$ (Eq.~\eqref{eq:AllOrder:NGLs}) and $\C(t)$ (Eq.~\eqref{eq:AllOrder:CLs}) for $k_t$ and C/A jet algorithms with $R = 0.7$. Solid lines: $k_t$; dashed lines: C/A. The black line shows the Monte Carlo output for the $k_t$ algorithm.}
\label{fig:CA:St-CT_R7}
\end{figure}
In Fig.~\ref{fig:CA:St-CT_R7}, we plot the NGLs and CLs exponentials (Eqs.~\eqref{eq:AllOrder:NGLs} and \eqref{eq:AllOrder:CLs}) for $k_t$ (solid lines) and C/A (dashed lines) algorithms with jet radius $R = 0.7$.
The two-loop results are identical, as mentioned in Sec.~\ref{sec:fixed-order}, hence the curves coincide. The three- and four-loop C/A curves for $\cS(t)$ are slightly higher than those for $k_t$, indicating a reduction in the corresponding fixed-order NGLs coefficients, as observed previously (recall the NGLs exponent is negative). This is clearly visible in the right-hand side plots, which show the ratio of C/A to $k_t$ form factors. Note that the calculations are more reliable for smaller $t$ \cite{Khelifa-Kerfa:2015mma}, and including more higher-loop coefficients extends the range of $t$ over which the results are reliable.
For CLs, since the exponent is positive, smaller terms yield a smaller exponent. We observe that the C/A curves are slightly lower than those for $k_t$, again indicating that their corresponding fixed-order coefficients are smaller in magnitude. This is clearly supported by the ratio plot. Note that for $\C(t)$, the difference between the three- and four-loop results is minuscule for both $k_t$ and C/A, largely due to the small coefficient sizes and the $1/n!$ prefactor.

\begin{figure}
\centering
\includegraphics[scale=0.51]{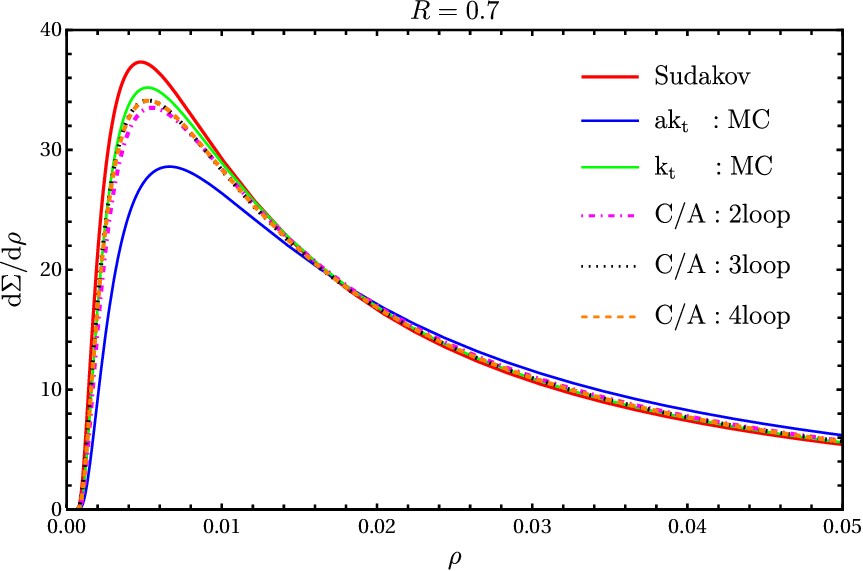}
\includegraphics[scale=0.51]{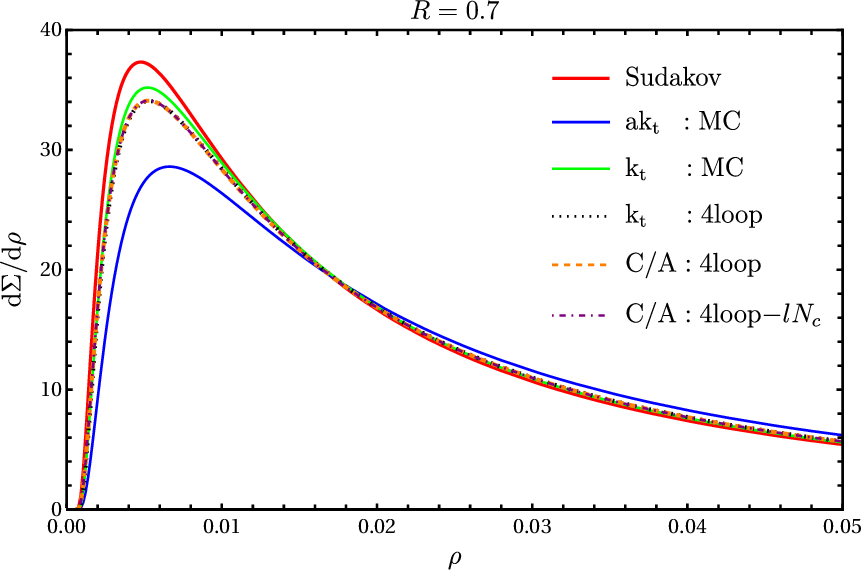}
\vskip 5pt
\includegraphics[scale=0.51]{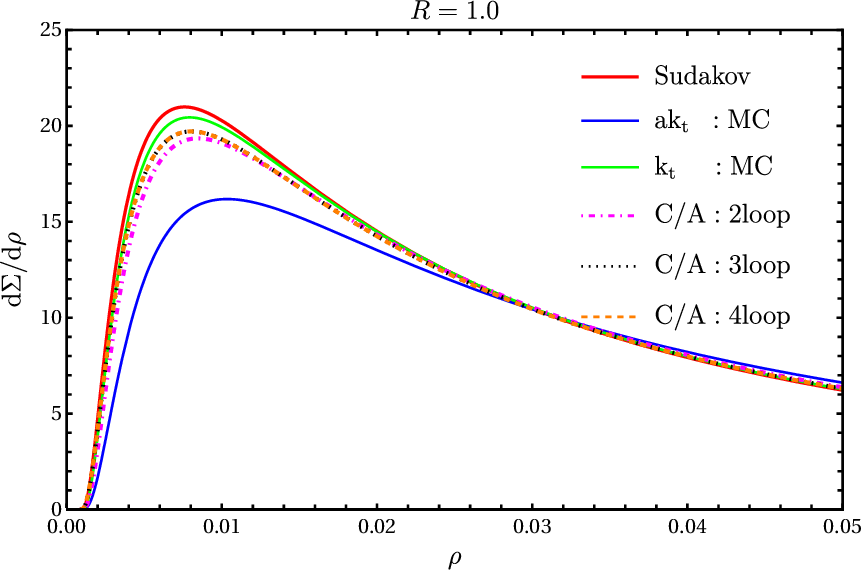}
\includegraphics[scale=0.51]{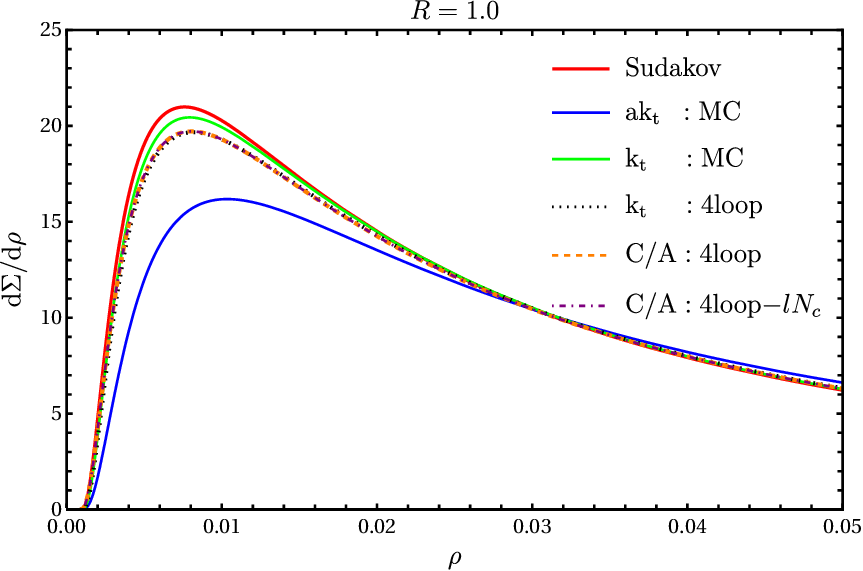}
\caption{The full NLL-resummed form factor of the jet mass observable for anti-$k_t$, $k_t$, and C/A jet algorithms with $R = 0.7$ and $R = 1.0$. The notation ``$lN_c$" denotes the large-$N_c$ limit.}
\label{fig:NLLFormFactor}
\end{figure}

In Fig.~\ref{fig:NLLFormFactor}, we plot the full NLL-resummed form factor \eqref{eq:AllOrder:Full} for three jet algorithms. For the anti-$k_t$ and $k_t$ algorithms, we use the MC parametrisations \eqref{eq:AllOorder:NGLs-MC} and \eqref{eq:AllOorder:CLs-MC} for the NGLs and CLs form factors, respectively, while for C/A clustering we use the exponentials \eqref{eq:AllOrder:NGLs} and \eqref{eq:AllOrder:CLs}.
The left-hand side plots compare the full form factor where the C/A NGLs and CLs exponentials include terms up to two, three, and four loops, for $R = 0.7$ and $R = 1.0$. The two-loop curves exhibit a shift in the Sudakov peak position and a reduction in its height of about $10\%$ for $R=0.7$ and $8\%$ for $R=1.0$. The three- and four-loop curves are nearly identical, again featuring a peak shift and height reduction of about $8\%$ for $R=0.7$ and $6\%$ for $R=1.0$.

The four-loop C/A results are plotted alongside the corresponding four-loop $k_t$ results, MC outputs for anti-$k_t$ and $k_t$, and the Sudakov form factor in the right-hand side plots of Fig.~\ref{fig:NLLFormFactor} for both $R$ values. The anti-$k_t$ curve exhibits the largest peak shift and Sudakov peak height reduction, reaching up to $23\%$. This is followed by the four-loop curves for both $k_t$ and C/A, with the C/A distribution slightly lower than that of $k_t$, showing reductions of up to $8\%$ and $6\%$ for $R = 0.7$ and $R=1.0$, respectively. The all-orders MC result for $k_t$ performs best, with Sudakov peak height reductions of only about $6\%$ and $3\%$ for the two $R$ values, respectively. Since the C/A four-loop distribution performs slightly better than the corresponding $k_t$ distribution in minimising the overall impact of NGLs and CLs, we expect the all-orders result to perform better than that of $k_t$.

It is worth noting that switching off the finite-$N_c$ corrections by setting $\CA = 2\CF$ in the four-loop NGLs coefficient for the C/A algorithm has no noticeable effect on the curves in Fig.~\ref{fig:NLLFormFactor}. This corresponds to the large-$N_c$ approximation, denoted by ``$lN_c$" in the plots.
Such a curve should closely approximate the result from a leading--colour, NLL-accurate dipole shower, such as \texttt{Panscales} \cite{vanBeekveld:2023ivn}.

\section{Conclusion}
\label{sec:conclusion}

In this paper, we have computed, for the first time in the literature, the full fixed-order distribution of a QCD non-global observable up to four loops at single-logarithmic accuracy in $e^+ e^-$ annihilation to dijets, where jets are defined using the Cambridge-Aachen sequential recombination algorithm. We have explicitly demonstrated that the structure of the C/A algorithm is considerably more complex, and the associated analytical calculations substantially more involved, than for its counterparts, the anti-$k_t$ and $k_t$ algorithms. The primary source of complexity stems from the fact that C/A involves no intrinsic ordering parameter. In other words, the distance between any pair of final-state particles is equally likely to be the smallest, in contrast to $k_t$, where clustering distances are weighted by the square of the particles' transverse momenta.

We have shown that while C/A is identical to $k_t$ at two loops, the algorithms begin to differ at higher orders. We have therefore presented detailed calculations of the full jet mass distribution at three and four loops at the single-logarithmic level (terms of the form $\as^n L^{n}$). These involve computing both CLs and NGLs contributions, which are particularly delicate to determine given the aforementioned complexity of the C/A algorithm for analytical calculations. In general, the C/A algorithm encompasses $k_t$, in the sense that the full contribution of the latter is contained within the former. Additionally, C/A introduces new corrections beyond the scope of the $k_t$ algorithm. These corrections are generally of opposite sign to the $k_t$ contributions for both NGLs and CLs, thereby reducing the overall impact of non-global logarithms to varying degrees.

The all-orders treatment of the NGLs and CLs contributions for the C/A algorithm was limited to comparisons with results from the anti-$k_t$ and $k_t$ algorithms, primarily due to the absence of all-orders analytical or numerical evaluations for C/A clustering. Overall, within the approximations and accuracy of this work, the C/A algorithm performs better than the other two in minimising the effect of non-global logarithms. Since non-global effects are notoriously difficult to compute—both at fixed order and to all orders, even at leading logarithmic accuracy—finding ways to mitigate them is highly desirable. The C/A algorithm appears to be a promising candidate for achieving this goal. Whether higher-loop calculations and/or all-orders resummation of C/A logarithms confirm this observation remains to be seen and will be the subject of future work.

It would be worthwhile to apply the current techniques and results to more complex QCD environments, exploring how the C/A algorithm compares to other jet algorithms in such multi-scale processes. Developing an integro-differential equation similar to that of Banfi, Marchesini, and Smye, or a computational code analogous to that of Dasgupta and Salam that incorporates C/A clustering, would constitute a major step toward a deeper and more insightful understanding of how complex jet algorithms reshape the impact of non-global logarithms.







\bibliography{Refs}


\end{document}